\documentclass[aps,pra,twocolumn,superscriptaddress,export]{revtex4-1} 
\usepackage[colorlinks=true,linkcolor=blue,citecolor=blue,urlcolor=blue]{hyperref}

\usepackage[utf8]{inputenc}
\usepackage[german, english]{babel}
\usepackage[dvipsnames]{xcolor}

\usepackage{times}

\usepackage{dcolumn}  
\usepackage{bm}        
\usepackage{bbold}
\usepackage{amsthm}
\usepackage{mathtools}
\usepackage{tikz}
\usetikzlibrary{shapes.geometric, arrows, shapes.multipart}
\usepackage{booktabs, tabularx}

\usepackage[export]{adjustbox}
\usepackage{titlesec}
\usepackage{amsfonts}
\usepackage{amssymb}
\usepackage{braket}

\usepackage{subfigure}

\newcommand{\diff}[1]{{{#1}}}

\newtheorem{Theorem}{Theorem}

\usepackage[normalem]{ulem}

\newtheorem{Lemma}{Lemma}

\begin{document}

\title{Universal measurement-based quantum computation in a one-dimensional architecture\\enabled by dual-unitary circuits}

\date{\today}

\author{David T. Stephen}
\affiliation{Department of Physics and Center for Theory of Quantum Matter, University of Colorado Boulder, Boulder, Colorado 80309 USA}
\affiliation{Department of Physics, California Institute of Technology, Pasadena, California 91125, USA}
\author{Wen Wei Ho}
\affiliation{Department of Physics, National University of Singapore, Singapore 117542}
\author{Tzu-Chieh Wei}
\affiliation{C. N. Yang Institute for Theoretical Physics, State University of New York at Stony Brook, Stony Brook, NY 11794-3840, USA}
\affiliation{Department of Physics and Astronomy,
State University of New York at Stony Brook, NY 11794-3800, USA}
\author{Robert Raussendorf}
\affiliation{Department of Physics and Astronomy, University of British Columbia, Vancouver, BC  V6T 1Z1, Canada}
\affiliation{Stewart Blusson Quantum Matter Institute, University of British Columbia, Vancouver, BC V6T 1Z4, Canada}
\author{Ruben Verresen}
\affiliation{Department of Physics, Harvard University, Cambridge, MA 02138, USA}

\begin{abstract}
{A powerful tool emerging from the study of many-body quantum dynamics is that of dual-unitary circuits, which are unitary even when read `sideways', i.e., along the spatial direction. 
Here, we show that this provides the ideal framework to understand and expand on the notion of measurement-based quantum computation (MBQC).
In particular, applying a dual-unitary circuit to a many-body state followed by appropriate measurements effectively implements quantum computation in the spatial direction.
We show how the dual-unitary dynamics generated by the dynamics of the paradigmatic one-dimensional kicked Ising chain with certain parameter choices generate resource states for universal deterministic MBQC.
Specifically, after $k$ time-steps, equivalent to a depth-$k$ quantum circuit, we obtain a resource state for universal MBQC on $\sim 3k/4$ encoded qubits.
\diff{Our protocol allows generic quantum circuits to be `rotated' in space-time and gives new ways to exchange between resources like qubit number and coherence time in quantum computers.}
Beyond the practical advantages, we also interpret {the dual-unitary evolution as generating an infinite sequence of new symmetry-protected topological phases \diff{with spatially modulated symmetries}, which gives a vast generalization of the well-studied one-dimensional cluster state} and shows that our protocol is robust to symmetry-respecting deformations.}
\end{abstract}

\maketitle

\textit{Introduction.}---Recent years have seen significant advances at the frontier of many-body quantum dynamics. A particularly fruitful approach has been to study   time-evolution induced by quantum circuits, minimal models of dynamics in which degrees of freedom are updated by local unitary gates.
Imposing structure on these gates lead to different classes of dynamics
---including Clifford~\cite{Gottesman1997}, matchgate~\cite{Valiant2002,Terhal2002,Jozsa2008}, and Haar random circuits~\cite{Nahum2017,Nahum2018,Khemani2018,Rakovszky2018,Suenderhauf2018,Chan2018,Keyserlingk2018,Fisher2022}---which allow for the the efficient computation of physical quantities while still capturing different interesting regimes of behavior.
A recent promising class is that of \textit{dual-unitary circuits}~\cite{Bertini2019,Akila2016,Gopalakrishnan2019,Bertini2020,Piroli2020,Suzuki2022,Kos2021,PhysRevLett.126.100603,Zhou2022,Borsi2022}, which are 
composed of gates that are not only unitary in the time-direction, as required by dynamics in closed quantum systems, but also unitary in the space-direction, upon exchanging the role of space and time. 
Despite this strong property, the class of dual-unitaries is broad and rich, capturing both integrable and chaotic systems~\cite{Bertini2019, PhysRevLett.126.100603}. 
The versatility of this approach has seen many applications,
allowing one to exactly compute spatio-temporal correlation functions~\cite{Bertini2019, PhysRevLett.126.100603}, spectral statistics~\cite{PhysRevLett.121.264101, Bertini2021_RMT_DU},  and entanglement dynamics~\cite{Piroli2020, PhysRevB.104.014301}, 
thereby providing deep insights into phenomena like quantum chaos, information scrambling and  thermalization.  

Intriguingly, the idea of regarding one spatial direction as an effective time-direction along which a circuit runs appears already in an older topic, namely that of \textit{measurement-based quantum computation} (MBQC)~\cite{Raussendorf2003,Raussendorf2012}. Here, the idea is that by creating an entangled many-body `resource' state using a finite-depth circuit and subsequently measuring the qubits, it is possible to effectively propagate quantum information through the spatial direction. The desired class of resource states is such that this spatial propagation is, indeed, unitary. Remarkably, there exist resource states in two spatial dimensions (2D) which are {\it universal}, meaning that this unitary evolution in the spatial direction can efficiently realize {\it any} quantum operation acting on {\it any} given number of encoded qubits~\cite{Raussendorf2001}. Decades of research has uncovered a plethora of such universal resources~\cite{Gross2007,Brennen2008,Kwek2012,Wei2018,Gachechiladze2019}, including a fault-tolerant protocol in 3D~\cite{Raussendorf2006}, but the full classification of universal resource states is still ongoing \cite{Prakash2015,Wang2017,Stephen2017,Miller2018,Stephen2019,Daniel2020,Herringer2022}. Both computation and fault-tolerance in MBQC have been realized in proof-of-principle experiments~\cite{Walther2005,Prevedel2007,Yao2012}.

\begin{figure}
	\centering
	\includegraphics[width=\linewidth]{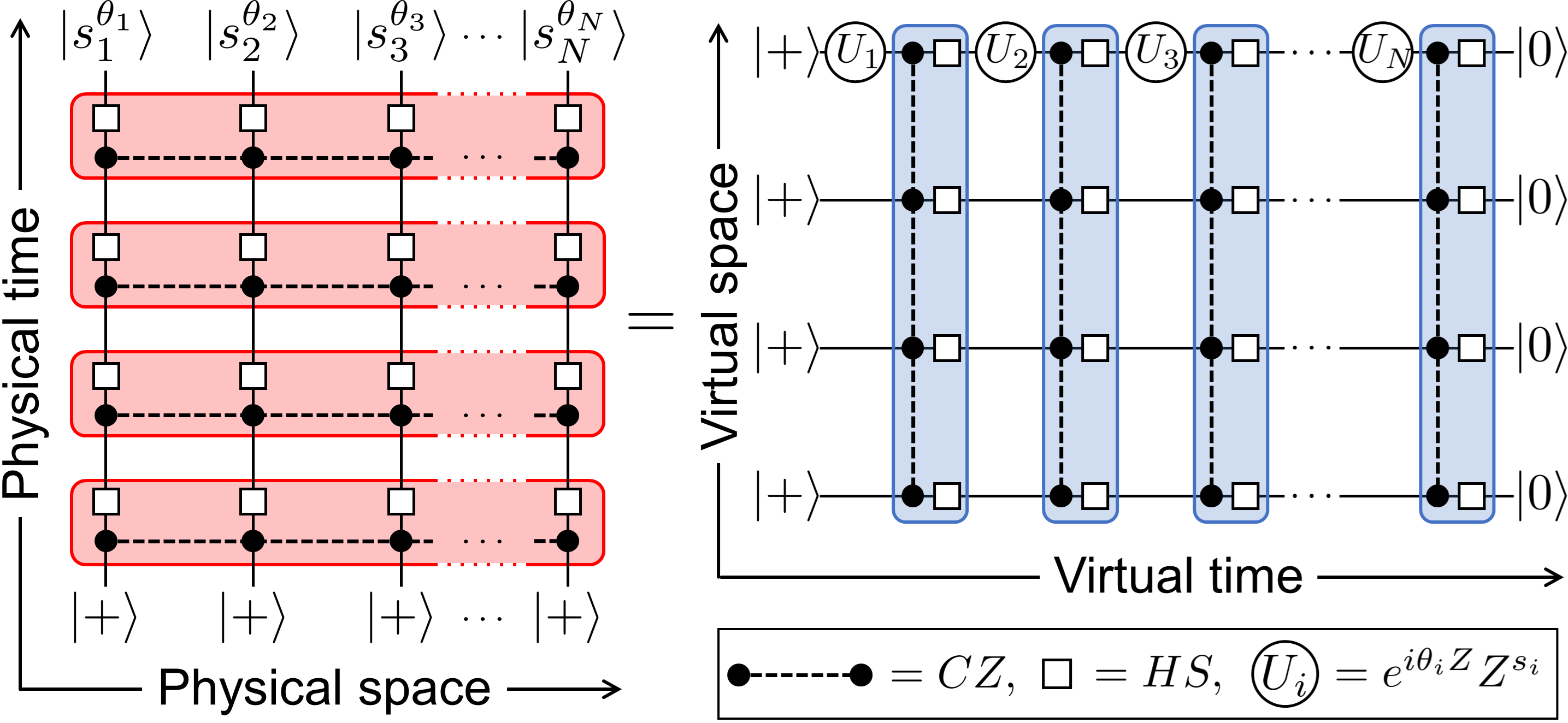}
	\caption{Graphical depiction of Eq.~\eqref{eq:dualunitary} for $k=4$. The red (blue) rounded box indicates one application of $T_N$ ($T_k$).
	In accordance with the convention of quantum circuits, the order of matrix multiplication in the right figure is from left to right. \diff{The left side describes the preparation and measurement of the resource state $|\psi_k\rangle$ while the right side is the simulated universal quantum circuit.}}
	\label{fig:dual_unitary}
\end{figure}

In this paper, we show how insights from dual-unitarity
can shed new light on MBQC, both at a conceptual and practical level. Namely, reading a dual-unitary circuit in the time direction describes the preparation of the resource state, while reading it in the spatial direction directly reveals the logical circuit induced by appropriate measurement of the resource state, as pictured in Fig.~\ref{fig:dual_unitary}. This provides an accessible alternative approach to MBQC beyond the traditional stabilizer~\cite{Raussendorf2003}, teleportation~\cite{Childs2005}, or matrix product state-based ~\cite{Gross2007} formalisms, and highlights how resource states can emerge naturally under certain classes of quantum many-body dynamics. These results also elevate dual-unitarity from an abstract computational tool to a concept with direct practical application. 

While Refs.~\onlinecite{Ho2022, Claeys2022emergentquantum,Ippoliti_Projected} explored how measurements can stochastically induce a universal gateset in the spatial direction of dual-unitary circuits, performing scalable quantum computation requires deterministic control without post-selection of measurement outcomes. Here, we show that,
by using dual-unitary circuits which are additionally chosen to be Clifford, it is possible to implement deterministic MBQC without sacrificing universality. Specifically, we show that a depth-$k$ dual-unitary circuit composed of repeated  applications of uniform single-site and nearest-neighbor Ising entangling gates on a chain prepares a resource state for MBQC on $k$ qubits. We then identify an unbounded sequence of $k$ for which the resulting unitary evolution that can be efficiently implemented via measurement is universal on at least $\sim 3k/4$ qubits. 
\diff{Practically, our protocol allows one to trade between space and time resources in quantum computers and to access deeper circuits than are possible with existing MBQC schemes using a given number of measurements, all while requiring only spatially-uniform nearest-neighbour and single-qubit operations in a 1D architecture.}

From another perspective, the dual-unitary circuits we consider can be viewed as ``infinite-order entanglers'' for symmetry-protected topological (SPT) orders~\cite{Pollmann2010,Chen2011,Schuch2011,Chen2013}. {That is, after $k$ time steps, the resulting state possesses 1D SPT order with a symmetry group and edge degeneracy that grow unboundedly with $k$}. On one hand, this shows that certain dual-unitary circuits provide a new way to generate infinite families of SPT order, which have proven very useful to the study of SPT order in the past~\cite{Fidkowski2011,Lahtinen2015,Verresen2017,Verresen2018,Jones19,Jones21a,Jones21,Balabanov21,Balabanov22}. On the other hand, this insight connects our results to the large literature on using SPT phases as resources for MBQC~\cite{Prakash2015,Wang2017,Stephen2017,Miller2018,Stephen2019,Daniel2020,Doherty2009,Bartlett2010,Miyake2010,Else2012,Miller2015,PoulsenNautrup2015,Miller2016,Wei2017,Raussendorf2019,Devakul2018}. In particular, it allows us to directly apply previous results~\cite{Else2012,Stephen2017,Stephen2019} to show that our MBQC schemes work not only using the fixed-point states generated by the dual-unitary evolution, but also any generic deformations thereof preserving certain symmetries.

\textit{Resource states from a dual-unitary circuit.}---
We consider $N$ qubits arranged on a line, denote Pauli operators including the identity as $I,X,Y,Z$, and write the eigenbasis of $Z$ as $|0\rangle,|1\rangle$. Dynamics are generated by the kicked Ising model, which is defined by the Floquet evolution
$U_F = e^{-i h \sum_{i=1}^N Y_i } e^{-i H_\text{Ising} \tau}$ where $H_\text{Ising} = J\sum_{i=1}^{N-1} Z_i Z_{i+1} + g \sum_{i=1}^N Z_i$ \footnote{{We note that the entanglement generation properties of $U_F$ were also previously studied in Refs.~\cite{Pal2018,Naik2019}.}}.
We fix $\tau=1$ and $J,h=\pm \pi/4$, since the evolution generated by $U_F$ is dual unitary only for these parameters. The parameter $g$ is chosen such that the evolution is also Clifford, meaning that any product of Pauli operators is mapped to another product of Pauli operators under conjugation by $U_F$. This important property will allow us compensate for random measurement outcomes and achieve deterministic computation. This property holds for $g=0$ and $g=\pm \pi/4$, but we focus on the latter and leave the former to the discussions at the end. 
$U_F$ is  then equivalent to the quantum circuit (up to irrelevant phases and Pauli operators
\footnote{Pauli operators and global phases will change the stabilizers $S_i^{(k)}$ and generators of rotations $O_k(\ell)$ only by a phase which affects neither the physics of the resource states nor their computational capability.}),
\begin{equation} \label{eq:qca}
   T_N = \prod_{i=1}^{N}H_iS_i \prod_{i=1}^{N-1} CZ_{i,i+1},
\end{equation}
where we define the gates $CZ = I-2|11\rangle \langle 11|$, $H = \frac{X + Z}{\sqrt{2}}$ and $S = \sqrt{Z} = \mathrm{diag}(1,i)$. We note that the single-qubit gate $HS$ cyclically permutes the three Pauli operators under conjugation.

The resource states $|\psi_k\rangle$ are defined by acting on an initial product state with the unitary circuit $k$ times, 
\begin{equation}
    |\psi_k\rangle = T_N^k\left(\bigotimes_{i=1}^N|+\rangle\right),
\end{equation}
where $|+\rangle = \frac{|0\rangle + |1\rangle}{\sqrt{2}}$. These states are pictured in Fig.~\ref{fig:dual_unitary}. Since $T_N$ is a Clifford circuit, the states $|\psi_k\rangle$ are stabilizer states that are uniquely defined by the equations $S^{(k)}_i|\psi_k\rangle=|\psi_k\rangle$, where $S^{(k)}_i = T_N^k X_i T_N^{k\dagger}$.
Equivalently, $|\psi_k\rangle$ is the unique ground state of the gapped Hamiltonian $H_k=-\sum_i S^{(k)}_i$. For example, we have $S^{(1)}_i = X_{i-1} Y_i X_{i+1}$ (with modifications  near the ends of the chain), so $|\psi_1\rangle$ is the 1D cluster state~\cite{Raussendorf2003}.
The 1D cluster state is a prototypical resource for MBQC on a single encoded qubit, and also a simple example of 1D SPT order~\cite{Son2012}. 

Now we will describe a protocol for MBQC using the states $|\psi_k\rangle$ as resource states. The equivalence of our measurement-based scheme to the traditional unitary gate-based model will follow directly from the dual-unitarity of $T_N^k$. In our protocol, each qubit in the chain is measured sequentially from left to right in a rotated basis $\{|0^{\theta}\rangle,|1^\theta\rangle\}$ defined by an angle $\theta$ where $|s^\theta\rangle = e^{-i\theta X}|s\rangle$.
The output of the quantum computation is determined by the probabilities of obtaining different measurement outcomes, which, according to the Born rule, are given by the inner products $|\langle s_1^{\theta_1},\dots,s_N^{\theta_N}|\psi_k\rangle|^2$ 
where each $s_i=0,1$.
To determine these overlaps, we utilize the dual-unitary property of the circuit $T_N^k$ to read it ``sideways'', which gives \cite{supplement},
\begin{equation} \label{eq:dualunitary}
    \langle s_1^{\theta_1},\dots,s_N^{\theta_N}|\psi_k\rangle = \langle R| U(\theta_N,s_N)\dots U(\theta_1,s_1)| L \rangle
\end{equation}
where we have defined the vectors $|L\rangle = \bigotimes_{i=1}^k |+\rangle_i$ and $| R \rangle = \bigotimes_{i=1}^k | 0\rangle_i$ and the unitary operator $U(\theta,s)=T_k e^{i\theta Z_1}Z_1^{s}$ where $T_k$ is as defined in Eq.~\eqref{eq:qca}. \diff{This equation, which is depicted in Fig.~\ref{fig:dual_unitary}, is the first step in proving universality of our protocol.}
It shows that the statistics arising from measuring the resource states $|\psi_k\rangle$ can be reinterpreted as describing a process in which $k$ ``virtual'' qubits are initialized in a state $|L\rangle$, evolved by unitaries $U(\theta,s)$, and then projected onto a final state $|R\rangle$. The evolution during this process depends on the choice of measurement bases defined by the angles $\theta_i$. Thus,
the dual-unitarity provides
a natural perspective on how measurement of physical qubits translates into controllable unitary evolution of the virtual qubits \footnote{We note that the picture developed here is equivalent to the picture of MBQC using tensor networks~\cite{Gross2007,Verstraete2004}, but in our scenario the tensor network structure emerges naturally from the quantum circuit diagrams, as opposed to introducing abstract tensors to represent some resource state.}.

\diff{The virtual computation described by the right-hand side of Eq.~\eqref{eq:dualunitary} currently has two issues. First, the unitary evolution depends on the measurement outcomes $s_i$, which are random. Second, the computation ends with a projection onto a fixed state $|R\rangle$ rather than a full projective measurement. It turns out that both issues are solved by adjusting future measurement bases depending on past measurement outcomes (as is common to all schemes of MBQC).}
We describe this in detail in Supplemental Material (SM) \cite{supplement}, and for the rest of the main text we always assume the outcome $|0^\theta\rangle$ is obtained. In short, the effect of obtaining the ``wrong'' measurement outcome $|1^\theta\rangle$ is to insert the byproduct operator $Z_1$ at that step in the computation. To deal with this unwanted operator, we imagine pushing it through to the end of the circuit. 
Importantly, because $T_k$ is Clifford, $Z_1$ will remain a product of Pauli operators as it is pushed through each layer of the circuit, which therefore only has two controlled effects:
First, depending on where the wrong outcome occurred, a subset of the rotation angles at later times will be flipped, $\theta_i\rightarrow -\theta_i$. This can be counteracted by flipping $\theta_i$ in the corresponding bases of future measurements. Second, once the byproduct operator is pushed to the end, it acts on $|R\rangle$ in such a way that $|R\rangle$ gets mapped onto a random product state in the $|0/1\rangle$-basis depending on the complete history of all measurement outcomes. Therefore, when accounting for the random measurement outcomes, repeating the protocol many times while recording the measurement statistics $|\langle s_1^{\theta_1},\dots,s_N^{\theta_N}|\psi_k\rangle|^2$
allows us to garner the measurement statistics $|\langle i_1,\dots , i_k|\phi_{\mathrm{out}}\rangle|^2$ for all $|i_j\rangle = |0/1\rangle$ where $|\phi_{\mathrm{out}}\rangle=U(\theta_N)\dots U(\theta_1)|L\rangle$ with $U(\theta_i)=T_k e^{i\theta_i Z_1}$ is the output state of the virtual computation. This constitutes a complete scheme of quantum computation, where we initialize a quantum register in a known state $|L\rangle$, perform deterministic unitary evolution on it, and read-out the final output state in a fixed basis. 

\begin{figure}
	\centering
	\includegraphics[width=\linewidth]{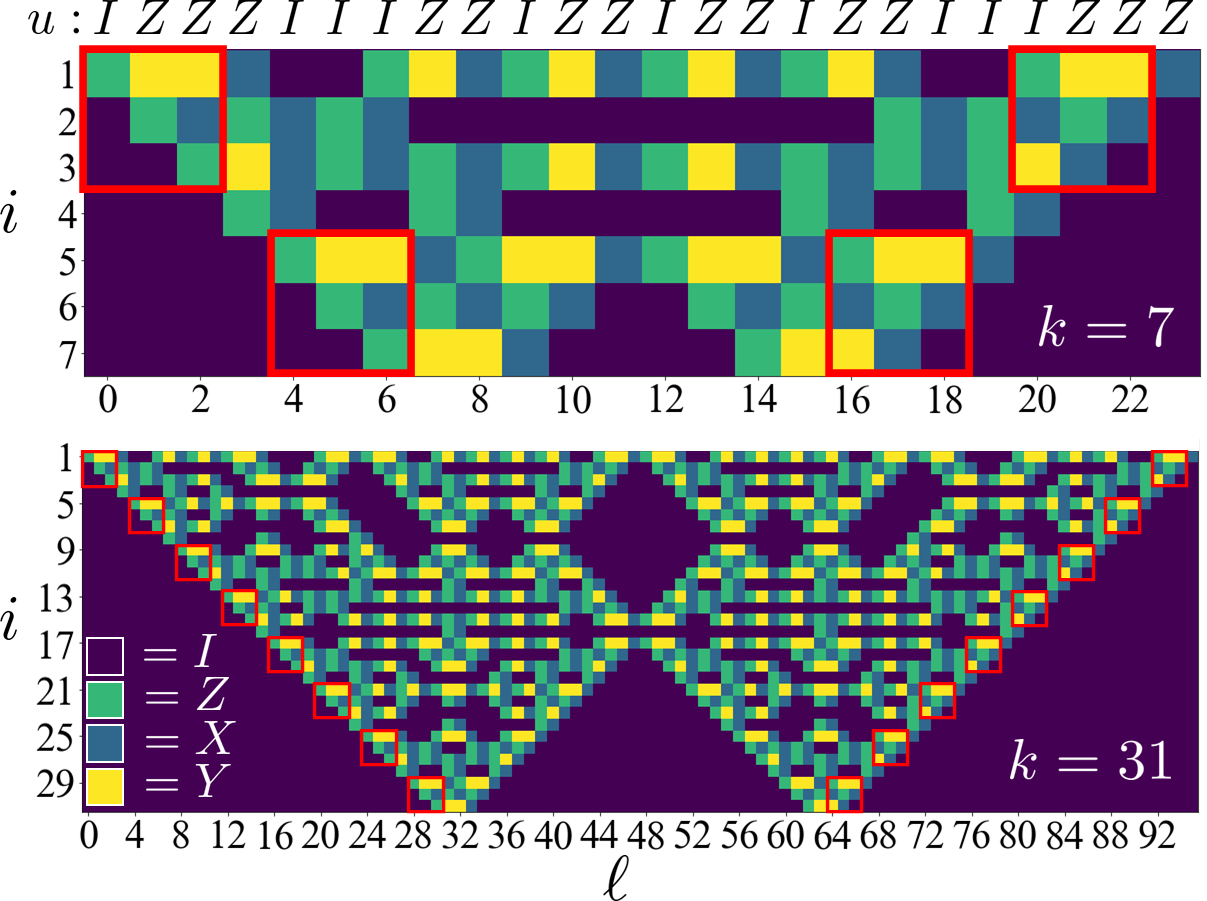} \\
	\vspace{2mm}
    \begin{tabular}[b]{|c|ccccccc|}\hline
      $k$ & 1 & 2 & 3 & 4 & 5 & 6 & 7  \\ \hline\hline
      $p_k$ & 3 & 4 & 12 & 10 & 24 & 18 & 24  \\ \hline
      dim($\mathcal{A}_k$) & 3 & 10 & 63 & 120 & 496 & 4095 & 8256  \\ \hline
      $\mathcal{A}_k$ & $su(2^k)$ & $sp(2^k)$ & $su(2^k)$ & $so(2^k)$ & $so(2^k)$ & $su(2^k)$ & $sp(2^k)$  \\ \hline
    \end{tabular}
	\caption{Top: Depiction of the operators $O_k(\ell)$ up to a phase for $k=7,31$ with respective periods $p_k=24,96$. Each column indexed by $\ell$ represents one product of Pauli operators acting on the $k$ virtual qubits which are indexed by $i=1,\dots,k$. The operator in one column is obtained from the previous via the local rules in Eq.~\eqref{eq:qca_rules}. The string of $I$ and $Z$ at the top indicates the repeating pattern found in the symmetry of $|\psi_k\rangle$ for $k=7$, where $Z$ coincides with $X$ and $Y$ operators in the top row of the space-time operator evolution. Bottom: Numerically determined computational power for small $k$. The lower row lists Lie algebras with dimensions consistent with the numerical calculations, where $su(n)$, $so(n)$, and $sp(n)$ denote the algebras of special unitary, orthogonal, and symplectic matrices.}
	\label{fig:qca_evo}
\end{figure}

\textit{Determining the set of gates.}---What remains is to determine which unitary circuits can be implemented using products of the unitaries $U(\theta)$. To understand these circuits, we  make the important observation that, since $T_k$ is unitary and Clifford, it has a finite period, meaning there is a smallest integer $p_k$ such that $T_k^{p_k} \propto I$. {The periods $p_k$ for $k\leq 7$ are given in Fig.~\ref{fig:qca_evo}.}
Now, consider breaking the computation into blocks of length $p_k$. The net effect of measuring all spins in one block is
\begin{equation}
    \prod_{\ell=0}^{p_k-1} U(\theta_\ell) = \prod_{\ell=0}^{p_k-1} T_k e^{i \theta_{\ell}Z_1} \propto \prod_{\ell=0}^{p_k-1} e^{i\theta_{\ell}O_k(\ell)},
\end{equation}
where $O_k(\ell) := T_k^{\ell\dagger} Z_1 T_k^\ell$. 
Therefore, the elementary gates in our scheme are $k$-qubit rotations generated by the operators $O_k(\ell)$, which are determined by the space-time evolution of $Z_1$ under conjugation by $T_k^\dagger$ a number $\ell$ times. Again, as $T_k$ is a Clifford circuit, these operators will all be $k$-qubit Pauli operators. The evolution is determined (up to a possible factor of $-1$) from the following local rules,
\begin{equation} \label{eq:qca_rules}
    T_k^\dagger X_iT_k = Z_i,\   
    T_k^\dagger Z_iT_k  = \begin{cases}
        Y_1Z_2 & i=1 \\
        Z_{i-1}Y_iZ_{i+1} & 1<i<k \\
        Z_{k-1} Y_k & i=k
    \end{cases}
\end{equation}
The space-time evolution of Pauli operators starting with $Z_1$ subject to these rules generates a fractal pattern that is pictured in Fig.~\ref{fig:qca_evo}. Let $\mathcal{O}_k$ denote the set of all $O_k(\ell)$ for $\ell = 0,\dots,p_k-1$.
For a small angles $d\theta$, we have $e^{id\theta P}e^{id\theta Q}\approx e^{id\theta(P+Q)}$ and $e^{id\theta P}e^{id\theta Q}e^{-id\theta P}e^{-id\theta Q} \approx e^{-(d\theta)^2[P,Q]}$ where $[P,Q]=PQ-QP$ for $P,Q\in\mathcal{O}_k$. Therefore, by concatenating our elementary gates, we can perform any rotation of the form $R=e^{i A}$ where $A$ is an element of the Lie algebra $\mathcal{A}_k$ generated by $\mathcal{O}_k$ using commutators. The set of such rotations is our set of implementable unitaries. 

Ideally, we desire $\mathcal{A}_k = su(2^k)$, i.e., $\mathcal{A}_k$ contains all $2^{2k}-1$ Pauli operators acting on $k$ qubits, which gives universal computation on all $k$ virtual qubits. At first glance, this seems impossible, since the only operations we use are $Z$ rotations of the first qubit and the application of a fixed $T_k$ to all qubits, so it is not clear how to selectively control a generic target qubit. However, we find that the persistent application of $T_k$ allows us to convert temporal control into spatial control. A similar concept was used in Ref.~\onlinecite{Raussendorf2005}. Indeed, looking at Fig.~\ref{fig:qca_evo}, we see that each operator $O_k(\ell)$ acts differently on different qubits. By judiciously combining these operators, we can selectively control all virtual qubits.

As a first investigation into the form of $\mathcal{A}_k$, we numerically generate the operators in $\mathcal{O}_k$ for small $k$ and repeatedly take commutators until no new operators are found. This is shown in Fig.~\ref{fig:qca_evo} for $k\leq 7$. In each case, we get an algebra with a dimension that scales exponentially with the number of qubits, but we only get the full $su(2^k)$ in certain cases. While the full algebra $\mathcal{A}_k$ appears to depend on $k$ in a complicated manner, we are able to prove the following lower bound on computational power:

\begin{Theorem} \label{res:universality}
{For every $k\geq 3$, let $m=\lfloor \frac{k+1}{4} \rfloor$. Then, the set of gates implementable in MBQC using the state $|\psi_k\rangle$ as a resource is universal on at least $3m$ qubits. That is, $su(2^{3m})\subset \mathcal{A}_k$. {If we further have $k=2^r-1$ for some $r\geq 0$, then the universal circuit model is guaranteed to be implemented with at most a linear overhead in $k$.}}
\end{Theorem}

\noindent
Therein, $\lfloor x \rfloor$ denotes the largest integer less than or equal to $x$. This is the main result of this work, as it means that the resource states $|\psi_k\rangle$ can be used for universal MBQC on $\sim 3k/4$ qubits.
The proof of this universality, given in SM \cite{supplement}, uses the self-similar fractal nature of the space-time evolution of the operators $O_k(\ell)$. Namely, we make use of repeating structures in this evolution indicated in Fig.~\ref{fig:qca_evo} to extend results for small values of $k$ to arbitrarily large values of $k$. To prove efficiency, we show that the period $p_k$---which essentially sets a clock speed for our computation since each elementary gate $e^{i\theta O_k(\ell)}$ can only be applied once per period---is linear in $k$ when $k=2^r-1$. 
{ 
Furthermore, the elementary gates in our scheme, namely the rotations $e^{i\theta O_k(\ell)}$, differ significantly from the standard gate set consisting of single-qubit rotations and nearest-neighbor two-qubit gates. 
Nevertheless, the proof of Theorem \ref{res:universality} shows how to efficiently construct any rotation of the form $e^{i\theta P}$ for an arbitrary Pauli string $P$---which includes the standard gate set---using a number of elementary gates $e^{i\theta O_k(\ell)}$ that is at most linear in $k$, such that our protocol can simulate the universal circuit model with polynomial overhead.
While reducing to the standard gate set is convenient to implement existing quantum algorithms, we note that it does not take full advantage of our gate set which, for example, also contains rotations that generate long-range entanglement in a single step (\textit{i.e.} those for which $O_k(\ell)$ is supported on a large fraction of the virtual qubits). 
}

\textit{Computational phases of matter.}---We have described a universal scheme of MBQC using the states $|\psi_k\rangle$ as resource states. It turns out that these states can also be interpreted as fixed-point states of certain 1D symmetry-protected topological (SPT) phases of matter \cite{Pollmann2010,Chen2011,Schuch2011}. To describe the SPT order, we first need to identify the symmetries. For this, we notice that Eq.~\eqref{eq:dualunitary} defines a matrix product state (MPS) representation of the wavefunction~\cite{Cirac2021}, from which the symmetries can be straightforwardly determined (see SM \cite{supplement}). We find that the symmetry group is generated by operators that form a string of $Z$ and $I$ which repeats along the chain with period $p_k$, \diff{similar to so-called spatially modulated symmetries \cite{Sala2021,Han2023}}. For $k=1$, the symmetry has the form $ZZIZZI\dots$ which repeats with a unit cell of size $p_1=3$.
In general, the repeating pattern mirrors the top row of the space-time evolution of the operators $O_k(\ell)$, such that the symmetry is also deeply linked to the dual-unitary structure, see Fig.~\ref{fig:qca_evo}. These symmetry operators and their translations generate the total symmetry group $\mathbb{Z}_2^{2k}$. 

The same MPS analysis also reveals the nature of the SPT order of the state $|\psi_k\rangle$ under the $\mathbb{Z}_2^{2k}$ symmetry. We find that the protected zero-energy edge mode that is characteristic of the SPT order has dimension $2^{k}$, which is the maximal possible value for this symmetry group. Because of this, the general results of Refs.~\onlinecite{Else2012,Stephen2017,Stephen2019} can be directly applied to our context to show that the MBQC protocol we have developed for the fixed-point states $|\psi_k\rangle$ works, with some modification, for any resource state coming from the same SPT phase.
Therefore, the ability to perform universal MBQC using single-site measurements is a property not only of the fine-tuned states $|\psi_k\rangle$, but also of the entire SPT phases of matter in which they reside.

\textit{Discussion.}---
\diff{ 
We have defined a protocol, enabled by dual-unitarity, to generate a new class of universal resource states for MBQC in a one-dimensional architecture using spatially-uniform controls.
Practically, our protocol represents a new way to embed quantum circuits in space-time. Namely, while the process implemented ``in the lab'' involves a circuit of depth-$k$ on $N$ physical qubits, the simulated circuit has depth-$N$ and $k$ qubits (up to constant factors) as in Fig.~\ref{fig:dual_unitary}. This allows one to trade between qubit number and coherence time in quantum computers, thereby making optimal use of available resources. Furthermore, simulating a depth-$N$ circuit on $k$ qubits using typical MBQC protocols would require measuring $Nk$ physical qubits \cite{Raussendorf2003}, so our resource states can access deeper circuits using the same number of measurements.
}

From a fundamental standpoint,
our results show that the dual unitary circuit $T_N^k$ can also be interpreted as an ``infinite-order SPT entangler''.
Namely, consecutive applications of $T_N$ to an initial product state generates an infinite sequence of 1D SPT orders with exponentially growing edge modes (when $N\rightarrow \infty$). {While all previously-defined circuits which generate SPT phases (i) have finite order such that $U^p=I$ for some $p$ independent of $N$ and (ii) generate phases with a fixed symmetry group $G$ \cite{Chen2013}, our circuit $T_N$ (i) has infinite order and (ii) generates SPT phases with a growing symmetry group $\mathbb{Z}_2^{2k}$.}
Our results therefore suggest the existence of a deep relationship between dual-unitary circuits, infinite-order SPT entanglers, and resource states for MBQC. We give a second example of this relationship in SM \cite{supplement} by replacing $HS\rightarrow H$ in $T_N$, corresponding to $g=0$ in the kicked Ising model. This circuit is also dual-unitary and generates an infinite family of 1D SPT ordered states that can be used for MBQC on $k$ virtual qubits.
However, in this case, the states are not universal resources since the set of gates implementable in MBQC generates an efficiently classically-simulable matchgate circuit.
This behaviour is likely fine-tuned, and we give a more general study in SM \cite{supplement}---where either $H$ or $HS$ is applied depending on the spatial location and time-step---which suggests the conjecture that generic dual-unitary Clifford circuits will generate resources for universal MBQC. We leave a deeper exploration into the relationships between these three concepts, as well as the induced classification of  MBQC resource states, for future work.

\begin{acknowledgments}
This work was initiated at the Aspen Center for Physics, which is supported by National Science Foundation grant PHY-1607611. RV is supported by the Harvard Quantum Initiative Postdoctoral Fellowship in Science and Engineering.
WWH acknowledges support from the National University of Singapore  start-up grants A-8000599-00-00 and A-8000599-01-00.
This work was
also partly supported by the Simons Collaboration on
Ultra-Quantum Matter, which is a grant from the Simons
Foundation (651440, DTS; 651440, RV). T.-C.W acknowledges partial support by the National Science Foundation  under Grant No. PHY 1915165 and  by the Materials Science and Engineering Divisions, Office of Basic Energy Sciences of the U.S. Department of Energy under Contract No. DESC0012704. RR is funded by NSERC and by USARO (W911NF2010013).
\end{acknowledgments}

\bibliography{biblio.bib}

\appendix

\section{Details of the dual unitary evolution} \label{sec:dual_unitary}

\begin{figure}
    \centering
    \includegraphics[width=\linewidth]{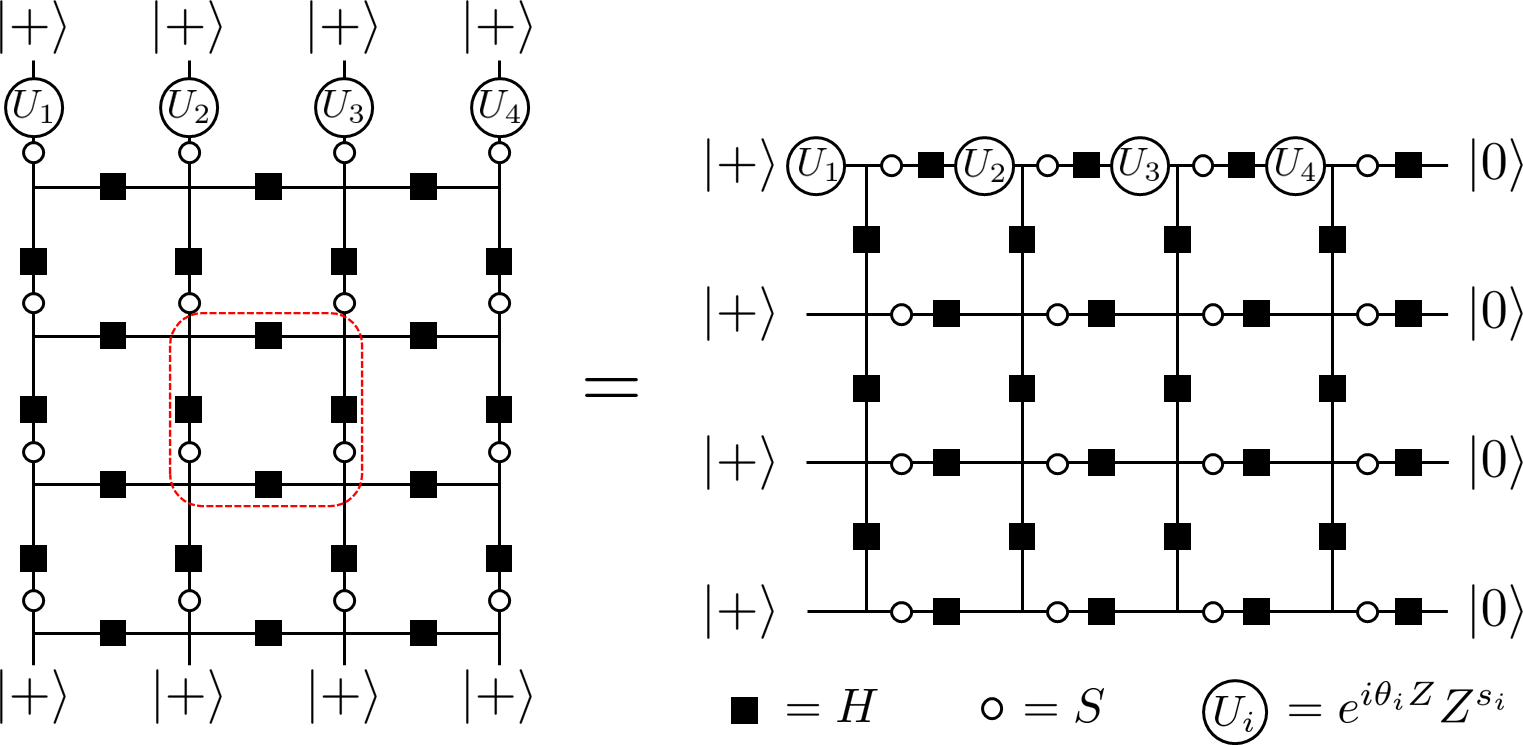}
    \caption{Tensor network representations of the left and right sides of Fig.~\ref{fig:dual_unitary} {The gates contained in the red box represent one dual-unitary gate $\mathcal{U} = CZ (HS\otimes HS) CZ$.}}
    \label{fig:dual_unitary_supp}
\end{figure}

In this section we give an explicit derivation of the equivalence between the two diagrams in Fig.~\ref{fig:dual_unitary}. To do this, we write each as a tensor network, and then perform simple tensor manipulations to demonstrate the equivalence of the two networks.
Such a diagrammatic approach was also utilized by Ref.~\cite{Ho2022} in their study of emergent state designs from partial measurements in the same circuit.
The main building block of these tensor networks is the $\delta$-tensor, indicated by a bare intersection of lines, 
\begin{equation}
    \includegraphics[scale=0.55,valign=c,raise=0.0cm]{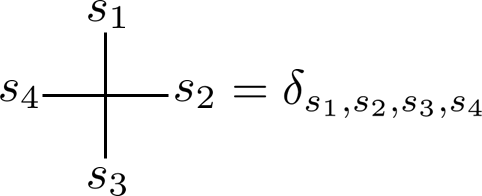}\, ,
\end{equation}
which enforces that all incoming indices take the same value in the $|0/1\rangle$ basis. The $\delta$-tensor transforms in the following simple way when one of the legs is contracted with the state $|+\rangle$,
\begin{equation} \label{eq:delta_tensor_contracted}
    \includegraphics[scale=0.275,valign=c,raise=0.0cm]{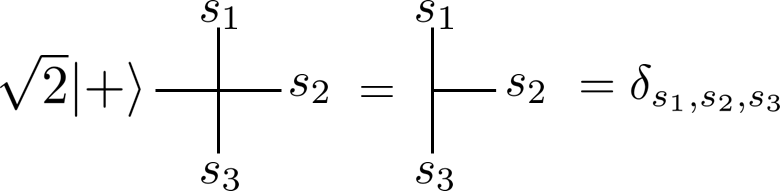}\, .
\end{equation}
The $CZ$ gate can be related to the Hadamard gate as $\langle ij |CZ|ij\rangle = \sqrt{2}\langle i|H|j\rangle$. This allows us to decompose the $CZ$ gate in terms of $\delta$-tensors and $H$ as follows,
\begin{equation}
    \includegraphics[scale=0.275,valign=c,raise=0.0cm]{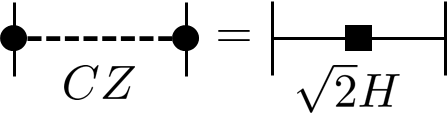}\, .
    \end{equation}
Finally, we note that we can write $\langle s^\theta| = \langle 0 |e^{i\theta X}X^s=\langle + | e^{i\theta Z}Z^s H$. Then, it is straightforward to see that the left-hand (right-hand) side of Fig.~\ref{fig:dual_unitary} can be expressed in terms of the tensor network depicted on the left-hand (right-hand) side of Fig.~\ref{fig:dual_unitary_supp}. The equivalence of these two tensor networks is then shown using the rule given in Eq.~\eqref{eq:delta_tensor_contracted}, and the fact that $S$ and $e^{i\theta Z}Z^s$ can freely move between legs of a $\delta$-tensor since they are diagonal in the $|0/1\rangle$ basis. This completes the proof of Eq.~\eqref{eq:dualunitary}.

{
Finally, we note that the circuit in Fig.~\ref{fig:dual_unitary} is not presented in the usual dual-unitary form which consists of a two-qubit gates that are themselves dual-unitary \cite{Bertini2019}. However, it can easily be cast into this form by grouping gates in into blocks defined as $\mathcal{U} = CZ (HS\otimes HS) CZ$, as shown in Fig.~\ref{fig:dual_unitary_supp}. It is easy to verify that $\mathcal{U}$ is indeed a dual-unitary gate. The whole circuit can then be expressed in terms of these dual-unitary gates, up to extra gates near the space-time boundaries of the circuit.
}

\section{Propagation of byproduct operators} \label{sec:byproduct_prop}

In the main text, we always assumed that a particular measurement outcome occurred for each measurement basis. In this section, we describe how to deal with the inherent randomness of measurement outcomes. In all cases described above, the effect of a ``wrong'' measurement outcome is to insert the operator $Z_1$ at that step of the computation. We show how to remove this by appropriately modifying the bases of all subsequent measurements, and discuss how this affects read-out at the end of computation. 

A sequence of gates in our protocol is described by a sequence of angles $\theta_i$ defining the measurement bases. Suppose we obtain the desired outcome $|0^{\theta_i}\rangle$ for each measurement. The total evolution is then described by the equation,
\begin{equation}
    \langle 0^{\theta_1},\dots,0^{\theta_N}|\psi_k\rangle = \langle R| U(\theta_N)\dots U(\theta_1)|L\rangle.
\end{equation}
where $U(\theta)=T_ke^{i\theta Z_1}$. Now suppose we obtain the wrong outcome $|1^{\theta_j}\rangle$ at some site $j$, but obtain the desired outcome everywhere else. This inserts the operator $Z_1$ at the $j$-th step of computation,
\begin{align}
    &\langle 0^{\theta_1},\dots,1^{\theta_j},\dots , 0^{\theta_N}|\psi_k\rangle \nonumber \\
    &=\langle R| U(\theta_N)\dots U(\theta_j)Z_1\dots U(\theta_1)|L\rangle \\
    &=\langle R| T_k^{N-j+1}(Z_1) \widetilde{U}(\theta_N)\dots \widetilde{U}(\theta_{j+1})U(\theta_j)\dots U(\theta_1)|L\rangle \nonumber
\end{align}
where we use the shorthand notation $T_k^l(\cdot)=T_k^l\cdot T_k^{l\dagger}$. In the second line, we have commuted the unwanted operator $Z_1$ to the end of the computation by writing $U(\theta_i)\mapsto \widetilde{U}(\theta_i)$ for all $j+1\leq i\leq N$ where,
\begin{equation}
    \widetilde{U}(\theta_i) = T_k^{i-j+1}(Z_1)^\dagger U(\theta_i)T_k^{i-j}(Z_1).
\end{equation}
Now we observe that $U(\theta_i)T_k^{i-j}(Z_1) = T_k^{i-j+1}(Z_1) U(\pm\theta_i)$ where the sign is $-1$ only if $T^{i-j}(Z_1)$ acts as $X$ or $Y$ on the first logical qubit, in which case it will anti-commute with $Z_1$, thereby flipping $e^{i\theta_i Z_1}\rightarrow e^{-i\theta_i Z_1}$. Then we have $\widetilde{U}(\theta_i)=U(\pm\theta_i)$.
More precisely, let $\xi=(\xi^X;\xi^Z)$ be a $2k$-component binary vector such that Pauli operators are written as $V(\xi) =\bigotimes_{i=1}^k X_i^{\xi^X_i}Z_i^{\xi^Z_i}$, and define the matrices $t_k$ acting on these vectors by $V(t_k\xi):=T_k V(\xi)T_k^\dagger$. Then, we have,
\begin{equation}
    \widetilde{U}(\theta_i)=U((-1)^{(t^{i-j}\hat{Z_1})^X_1}\theta_i),
\end{equation}
where $\hat{Z_1}=(0,\dots,0;1,0,\dots,0)$ is the vector representing the Pauli $Z_1$ and $(t_k^{i-j}\xi)^X_1$ denotes the first entry of the $X$-component of the vector $t_k^{i-j}\xi$.

Therefore, inserting the operator $Z_1$ at step $j$ has two effects. First, it flips the rotation angle of some future rotations in a pattern that is determined by $T_k$. This effect is counteracted by flipping the angle in the corresponding measurement basis, \textit{i.e.} sharing a classical bit of information $(t_k^{i-j}\hat{Z_1})^X_1$ with all sites $i>j$ which says whether this measurement angle should be flipped or not. These bits are then updated every time a new measurement outcome is obtained. This modification of future measurement bases in terms of past measurement outcomes can also be interpreted as applying symmetry transformations to the chain after each measurement outcome~\cite{Else2012a}. These symmetries are described in Section~\ref{sec:spt}.

The second effect of inserting $Z_1$ is to modify the boundary vector $\langle R|\mapsto \langle R| T_k^{N-j+1}(Z_1)$. This has a positive effect: since $\langle R|$ is initially in the state $\langle 0,\dots 0|$, this action will flip it (up to a sign) to another product state in the $Z$-basis that depends on $j$. In Section~\ref{sec:spt}, it is proven that the operators $T_k^{\ell}(Z_1)$ generate all Pauli operators under multiplication. Therefore, for all product states in the $|0/1\rangle$ basis, there exists a sequence of measurement outcomes that will map $|R\rangle$ to that state. Because of this, repeating our computation many times allows us to estimate $\langle i_1,\dots , i_k|\phi_{\mathrm{out}}\rangle$ for all $|i_j\rangle = |0/1\rangle$ where $|\phi_{\mathrm{out}}\rangle=U(\theta_N)\dots U(\theta_1)|L\rangle$. 

\section{Proof of Theorem~1} \label{sec:universality_proof}

Here we give the proof of Theorem~1 by explicitly constructing a universal gate set. The proof proceeds in a number of technical steps, but the main ideas come from Lemmas~\ref{lemma1} and~\ref{lemma2}. Throughout this section, we let $\mathcal{P}_k$ denote the set of all Pauli operators on $k$ qubits, not including the identity operator. Given a set $S$ of Paulis, we let $\mathcal{A}_S$ denote the Lie algebra generated by $S$ under commutation. The first part of the Theorem applies to all values of $k$, but we first fix $k=4m-1$ with $m>0$ and explain how to extend to general $k$ at the end.

\begin{Lemma} \label{lemma1}
Consider a set of Pauli operators $\mathcal{S}$ whose elements have the form $P_1\otimes P_2$ where $P_\alpha \in \mathcal{P}_{k_\alpha}$ with $k_\alpha>2$. Suppose that (i) there is a set of operators $S_1 = \{P_i\otimes I: i=1,\dots,M\} \subset \mathcal{S}$ which generate all operators $P\otimes I$ for $P\in \mathcal{P}_{k_1}$ under commutation and (ii) there is a set of operators  $S_2 = \{A_i\otimes B_i: i=1,\dots, M'\}\subset \mathcal{S}$ such that $A_i\neq I$ and the operators $B_i$ generate all of $\mathcal{P}_{k_2}$ under multiplication. Then the Lie algebra $\mathcal{A}_S$ contains all of $\mathcal{P}_{k_1+k_2}$, \textit{i.e.} $\mathcal{A}_\mathcal{S}=su(2^{k_1+k_2})$.
\end{Lemma}

\begin{figure}
    \centering
    \includegraphics[width=\linewidth]{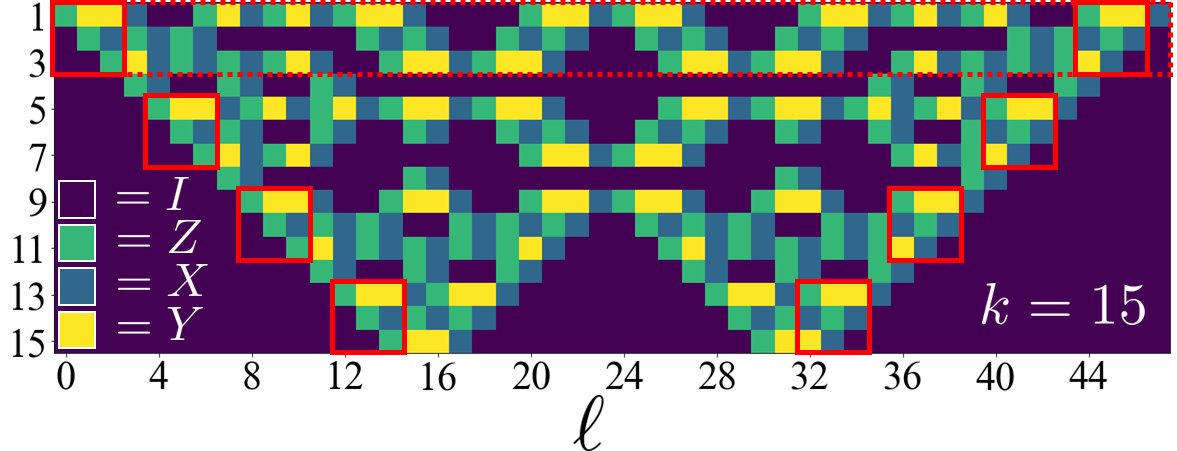}
    \caption{Depiction of the operators $O_k(\ell)$ for $k=15$. The solid boxes indicate repeating structures used in the proof of Theorem~1. The dotted box indicates the first three rows, which we observe are non-empty for every column as required for Lemma 
    \ref{lemma1}.}
    \label{fig:qca_evo_app}
\end{figure}

To use the above lemma, we will need to modify the operators $O_k(\ell)$ defined in the main text. Let us introduce the shorthand notation $O_k(-\ell)=O_k(p_k-\ell)$. Decompose these operators as $O_k(\ell) =\bigotimes_{i} O_k(\ell,i)$ where $O_k(\ell,i)$ is a Pauli operator acting on the $i$-th logical qubit and $i=1,\dots, k$ (the coordinates $\ell,i$ specify a column and row of the space-time evolution pictured in Fig.~\ref{fig:qca_evo_app}). Now define the following modified operators,
\begin{equation} \label{eq:otilde}
    \widetilde{O}_k(\ell) = 
    \begin{cases}  
        \bigotimes_{ i \notin 4\mathbb{N}} O_k(\ell,i) & \text{if } O_{k}(\ell,j) =I \text{ or } X \quad \forall j\in 4\mathbb{N}\\
        0 & \text{else}
    \end{cases}
\end{equation}
Therein, $4\mathbb{N}$ denotes all strictly positive multiples of 4. Let $\widetilde{\mathcal{O}}_k$ denote the set of all $\widetilde{O}_k(\ell)$. The purpose of defining this set of operators is as follows. Suppose that the state of the logical qubits is such that the qubits indexed by $i=4j$ for $j\in\mathbb{N}$ are in the state $|+\rangle$. Then $O_k(\ell)$ acts the same on this state as $\widetilde{O}_k(\ell)$ if the first case in Eq.~\eqref{eq:otilde} is fulfilled since $X|+\rangle = |+\rangle$. If the second case in Eq.~\eqref{eq:otilde} is fulfilled, then $\widetilde{O}_k(\ell)$ will change the state of at least one of the $4j$-th qubits. To prevent this from happening, we remove such operators from $\widetilde{\mathcal{O}}_k$ by setting them to 0. So, if we freeze every $4j$-th qubit in the state $|+\rangle$, we can treat $\widetilde{O}_k(\ell)$ as our effective generators of rotation. As an example, if $O_k(\ell)=Z_3X_4Z_5$, then $\widetilde{O}_k(\ell)=Z_3Z_5$.

\begin{Lemma} \label{lemma2}
For every $k=4m-1$, define a set of operators $\mathcal{S}_k\subset \widetilde{\mathcal{O}}_k$ as,
\begin{equation}
    \mathcal{S}_k=\left\{\widetilde{O}_k(\ell): \ell\in\mathcal{I}_k\right\},
\end{equation}
where,
\begin{equation}
    \begin{aligned}
        \mathcal{I}_k=\left\{ -1\right\}\cup \bigcup_{j=0}^{m-1}\{&4j,4j+1,4j+2,\\
        -&4j-2,-4j-3,-4j-4 \}
    \end{aligned}
\end{equation}
Then $\mathcal{S}_k$ satisfies the conditions outlined in Lemma~\ref{lemma1} { with $k_1=3$ and $k_2=3(m-1)$}, so $\mathcal{A}_{\mathcal{S}_k}=su(2^{3m})$.
\end{Lemma}

\noindent
In Fig.~\ref{fig:qca_evo_app}, the set $\mathcal{I}_k$ corresponds to the columns containing red boxes (along with $\ell=-1$). 

\vspace{5mm}
\noindent
\textit{Proof of Theorem~1.}
Combining Lemmas~\ref{lemma1} and~\ref{lemma2} shows that if the logical qubits indexed by $4j$ for $j>0$ are initialized in the state $|+\rangle$, then we have a universal set of gates on the remaining $3(k+1)/4 = 3m$ logical qubits. Our initial state $|L\rangle$ satisfies this since all qubits start in the state $|+\rangle$. So we have shown that $su(2^{3m})\subset \mathcal{A}_k$ when $k=4m-1$. To extend this to all values of $k$, let $q$ be the largest number $4m-1$ that is less than or equal to $k$. Then, since $\widetilde{O}_q(\ell)\equiv \widetilde{O}_k(\ell)$ for $|\ell|\leq k$, where we use $\equiv$ to denote equivalence up to padding with identity operators, we can simply apply Lemma~\ref{lemma2} with $k\rightarrow q$ to get $su(2^{3m})\subset \mathcal{A}_k$ where $m=\lfloor \frac{k+1}{4} \rfloor$.

{
We have proven that our gate set generates all unitary operators acting on the $3m$ logical qubits. To prove the second part of Theorem~1, we need to discuss efficiency. To do this, we will show that the universal circuit model, \textit{i.e.} the standard gate set, can be implemented with at most a linear overhead using our gate set. The standard gate set corresponds to the operators $e^{i\theta Z_j}$, $e^{i\theta X_j}$, and $e^{i\theta Z_jZ_{j+1}}$ where $j$ runs across all logical qubits. Note that we can perform rotations about an axis described by the commutator of two Paulis without requiring small angles using the equation $e^{i\frac{\pi}{4}P}e^{i\theta Q}e^{-i\frac{\pi}{4}P} = e^{-\theta [P,Q]/2}$ which holds for any pair $P,Q$ of anticommuting Paulis. Therefore, any sequence of commutators of the Paulis $\widetilde{O}_k(\ell)$ which equals a target Pauli $P$ directly translates to a sequence of gates to exactly realize the rotation $e^{i\theta P}$ with approximately double the length of the sequence of commutators. The proofs of Lemmas \ref{lemma1} and \ref{lemma2} show that such a sequence of commutators of the Paulis $\widetilde{O}_k(\ell)$ equalling $P$ can always be found. In fact, the proofs are constructive, and they give a procedure to derive such a sequence whose length grows at most linearly with $k$. The computational complexity of the procedure for deriving this sequence is also polynomial in $k$. Therefore, our gate set can simulate the standard gate set with a gate overhead that is at most linear in $k$, and the algorithm to compile into the standard gate set has polynomial complexity in $k$.
}

The other aspect which needs to be considered is that we are restricted to performing the rotation generated by $O_k(\ell)$ only once per ``clock cycle'' of length $p_k$ for each $\ell$. In principle, if $p_k$ grew exponentially large in $k$ (which is the case for certain values of $k$), this restriction could mean that we cannot efficiently generate arbitrary gates. Luckily, we have the following Lemma,
\begin{Lemma} \label{lemma3}
 For $k=2^r-1$, we have $p_k=3k+3$.
\end{Lemma}
\noindent
Therefore, $k$ can always be chosen such there is at most a linear overhead due to the ``clock cycle'' of the computational model. \hfill$\Box$

This linear overhead assumes that we perform only one rotation per clock cycle. In practice, it is likely that a given circuit can be compiled to our gate set in such a way that multiple rotations are performed per clock cycle, so this overhead will not saturate this linear bound. Even in cases where the period is exponentially large in $k$, the overhead is accompanied by a much larger set of elementary gates that can be done in a single time-step. This means we need to spend fewer resources
generating higher-order commutators of the $O_k(\ell)$ to get arbitrary elements of the Lie algebra. So it may in certain cases be advantageous to choose $k$ such that $p_k$ is exponentially large.

\vspace{5mm}
\noindent
\textit{Proof of Lemma \ref{lemma1}.}
Let us introduce a function $\Lambda$ such that $\Lambda(P,Q)=1$ if the Pauli operators $P,Q$ commute and $\Lambda(P,Q)=-1$ if they anticommute. Throughout this proof, we ignore all constant factors that may appear in front of Pauli operators, and write equality to mean equality up to a constant (for example, we may write $[X,Y]=Z$). Then, a commutator of two Pauli operators can be taken to be their product if they anticommute, and 0 if they commute. By definition, the set of Paulis $\mathcal{S}$ generates all Paulis under multiplication. The difficulty is to show that they also generate all Paulis under commutation. 

First we show that, if $P_1\otimes Q_1,P_2\otimes Q_2\in\mathcal{A}_\mathcal{S}$ with $P_1,P_2\neq I$ and $Q_1\neq Q_2$, then there exists $P_3\neq I$ such that $P_3\otimes Q_1Q_2\in \mathcal{A}_\mathcal{S}$. That is, we can effectively multiply $Q_1$ and $Q_2$, regardless of whether $P_1\otimes Q_1$ and $P_2\otimes Q_2$ commute or not. Let $\Lambda(Q_1,Q_2)=s$ and identify a Pauli $P_1'$ such that $P_1'\neq P_2$ and $\Lambda(P_1',P_2)=-s$; such an operator $P_1'$ always exists since $P_2$ always (anti)commutes with more than one other Pauli if $k_1>1$ ($P_1$ may already satisfy this, in which case the next step can be skipped). Now we want to show that $P_1'\otimes Q_1\in\mathcal{A}_\mathcal{S}$. If $\Lambda(P_1,P_1')=-1$, then, since $P_1P_1'\otimes I$ is in $\mathcal{S}$ by assumption, we have $[P_1\otimes Q_1,P_1P_1'\otimes I]=P_1'\otimes Q_1\in \mathcal{A}_\mathcal{S}$. If $\Lambda(P_1,P_1')=1$, we can always find $P_1''$ such that $\Lambda(P_1,P_1'')=\Lambda(P_1',P_1'')=-1$. Then $P_1''\otimes I$ and $P_1P_1'P_1''\otimes I$ are in $\mathcal{S}$ by assumption, so $[[P_1\otimes Q_1,P_1''\otimes I],P_1P_1'P_1''\otimes I]=P_1'\otimes Q_1\in \mathcal{A}_\mathcal{S}$. Finally, since $\Lambda(P_1'\otimes Q_1,P_2\otimes Q_2)=-1$ by construction, we have $[P_1'\otimes Q_1,P_2\otimes Q_2]=P_1'P_2\otimes Q_1Q_2\in\mathcal{A}_\mathcal{S}$ and $P_1'P_2\neq I$.

By assumption, the operators $B_i$ defined in the Lemma generate all of $\mathcal{P}_{k_2}$ via multiplication. Using the procedure described above, we can effectively multiply the operators $B_i$ using commutators. In this way, for all $Q\in \mathcal{P}_{k_2}$, there exists $P_Q\neq I$ such that $P_Q\otimes Q\in \mathcal{A}_\mathcal{S}$. By taking commutators of these elements with elements of $S_1$, any element of the form $P\otimes Q$ where $P,Q\neq I$ can be generated. Finally, we have $[P\otimes Q_1,P\otimes Q_2]=I\otimes [Q_1,Q_2]\in\mathcal{A}_\mathcal{S}$. Since every $Q\neq I$ can be written as $Q=Q_1Q_2$ for some $Q_1,Q_2$ with $\Lambda(Q_1,Q_1)=-1$, we have $Q=[Q_1,Q_2]$, so we generate all elements of the form $I\otimes Q$, which gives $\mathcal{A}_\mathcal{S}=\mathcal{P}_{k_1+k_2}=su(2^{k_1+k_2})$. \hfill $\Box$

\begin{figure}
    \centering
    \includegraphics[width=0.8\linewidth]{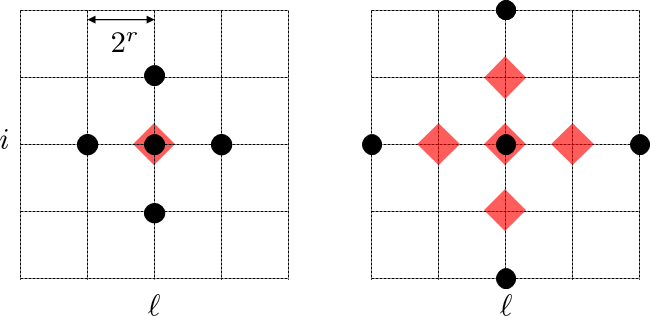}
    \caption{Illustration of the recurrence relation described by Eq.~\eqref{eq:recurrencer}. On the left, we use black dots to represent the bits $a_k(\ell,i)$. The five marked bits, separated by a scale $2^r$, satisfy a relation which says they must sum to 0 mod 2, as stated in Eq.~\eqref{eq:recurrence} for the case $r=0$. We use a diamond to indicate the center of this five-body relation. By summing five of these relations, with positions marked by the five diamonds, some of the bits involved in the relations cancel pairwise, and we retrieve the same recurrence relation over a distance $2^{r+1}$.}
    \label{fig:recurrence}
\end{figure}

\vspace{5mm}
\noindent
\textit{Proof of Lemma \ref{lemma2}.} Throughout this proof, we will make extensive use of the self-similar fractal nature of the space-time diagrams that describe the operators $O_k(\ell)$, as displayed in Fig.~\ref{fig:qca_evo} of the main text. To capture these patterns, we define the bits $a_{k}(\ell,i)$ and $b_{k}(\ell,i)$ such that $O_k(\ell,i)=Z_i^{a_k(\ell,i)}X_i^{b_k(\ell,i)}$, up to a phase. The following recurrence relation can be derived from Eq.~\eqref{eq:qca_rules},
\begin{equation} \label{eq:recurrence}
\begin{aligned}
    a_k(\ell+1,i)&\equiv a_k(\ell,i-1) + a_k(\ell,i) + a_k(\ell,i+1)\\ 
    & + a_k(\ell-1,i),
    \quad\text{for}\quad 1\leq i\leq k
\end{aligned}
\end{equation}
where $\equiv$ denotes equivalence modulo 2. This relation is subject to periodic boundary conditions in the $\ell$-direction such that $a_k(-\ell,i)=a_k(p_i-\ell,i)$, and open boundaries in the $i$-direction such that $a_k(\ell,0)=a_k(\ell,k+1)=0$. The same relation holds for $b_k(\ell,i)$. Additionally, we have $b_k(\ell,i)=a_k(\ell+1,i)$. Concatenating these relations leads to a hierarchy of relations of the following form,
\begin{equation} \label{eq:recurrencer}
\begin{aligned}
    a_k(\ell+2^r,i)&\equiv a_k(\ell,i-2^r) + a_k(\ell,i) + a_k(\ell,i+2^r)\\ 
    & + a_k(\ell-2^r,i),
    \quad\text{for}\quad 1\leq i\leq k
\end{aligned} 
\end{equation}
where $r\in\mathbb{N}$ is a number such that $2^r<k$. See Fig.~\ref{fig:recurrence} for a pictoral proof by induction of this equation. The boundary conditions satisfy $a_k(\ell,-i)=a_k(\ell,i-2)$ and $a_k(\ell,k+i)=a_k(\ell,k+2-i)$ for $1\leq i\leq k$. That is, the boundary conditions ``reflect'' indices outside of the range $1\leq i \leq k$ back into this range, as described for a similar scenario in Ref.~\onlinecite{Raussendorf2005} and discussed in more detail in the proof of Lemma~\ref{lemma3}. This recurrence relation captures the self-similarity of the operator evolution.

As a first step, we need to understand the form of the operators $\widetilde{O}_k(\ell)$ which appear in the definition of $\mathcal{S}_k$. We can show that none of them are equal to 0, meaning that the corresponding operators $O_k(\ell)\in\mathcal{O}_k$ satisfy $O_k(\ell,j) =I \text{ or } X \quad \forall j\in 4\mathbb{N}$. Using Eq.~\eqref{eq:recurrencer} with $2^r=4$, the initial conditions specified by our problem, and the fact that $k=4m-1$, we can show, for $m\in\mathbb{N}$,
\begin{equation} \label{eq:4throw}
    O_k(\ell,4m) = 
    \begin{cases}
        I \text{ or } X & \ell \in 4\mathbb{N} \\
        I & \ell \in 4\mathbb{N} + 1 \\
        I & \ell \in 4\mathbb{N} + 2 \\
        I \text{ or } Z & \ell \in 4\mathbb{N} + 3 \\
    \end{cases}.
\end{equation}
Since $\mathcal{S}_k$ contains only operators with $\ell\notin 4\mathbb{N} + 3$, this establishes the claim.

{

We observe an important repeating structure in the operators $\widetilde{O}_k(\ell)$. Namely, we have $O_k(\pm(\ell+ 4),i+4) = O_k(\pm\ell, i)$ for $\ell=0,\dots,k-4$ and $i=\ell,\dots,\ell+3$, as depicted by the red boxes in Fig.~\ref{fig:qca_evo_app}. This is because, for the specified values of $\ell$ and $i$, the relation in Eq.~\eqref{eq:recurrencer} for $2^r=4$ says that $a_k(\pm(\ell+ 4),i)=a_k(\pm\ell,i+ 4)$ (and similarly for $b_j(\ell,i)$) since all other terms in the relation are 0. 

The properties of the repeating operators allow us to prove the conditions of Lemma \ref{lemma1}. The first condition requires us to identify a subset of operators $S_1$ which generate all Paulis on the first three qubits under commutation. This is satisfied by taking $S_1=\left\{\widetilde{O}_k(\ell):\ell=0,1,2,-1,-2,-3,-4\right\}$. These operators all act only on the first three qubits, and we can confirm by a straightforward brute force calculation that they generate all Pauli operators under commutation. Note that, if we were to remove $\widetilde{O}_k(-1)$ from $S_1$, then the resulting set of six operators still generates all Paulis by \textit{multiplication}, but they no longer generate all Paulis under commutation. This is important for the following step.

The set $S_2$ specified in Lemma \ref{lemma1} is given by the rest of the operators in $\mathcal{S}_k$, \textit{i.e.} $S_2 = \mathcal{S}_k \setminus S_1$. Following the statement of Lemma \ref{lemma1}, we split all operators in $S_2$ as $\widetilde{O}_k(\ell) = A_\ell \otimes B_\ell$, where $A_\ell$ acts on the first three qubits ($i=1,2,3$), and $B_\ell$ acts on the rest ($i>3$). We need to show that (a) $A_\ell\neq I$ for all $\ell$ and (b) the operators $B_\ell$ generate all Paulis on qubits $i>3$ under multiplication. Condition (b) follows from the above observation that the repeating operators in the red boxes generate all Paulis on the respective three logical qubits under multiplication. Namely, for every $j\geq 1$, the operators $B_\ell$ for $\ell=4j,4j+1,4j+2,-4j-2,-4j-3,-4j-4$ generate all Paulis on qubits $i=4j+1,4j+2,4j+3$ under multiplication, up to some action on qubits $i<4j+1$. Therefore, the $B_\ell$ altogether generate all operators on the logical qubits with index $i>3$ under multiplication, as required.

Finally, we need to show that the operators $O_k(\ell)$ always act non-trivially on at least one of the qubits $i=1,2,3$, which implies condition (a) stated above. This can be readily observed to hold for the space-time evolutions pictured in Fig.~\ref{fig:qca_evo_app} and Fig.~\ref{fig:qca_evo}. The proof that this holds for arbitrary $k$ is somewhat tedious and not very insightful, and we therefore do not include it in full. In essence, the proof looks at the possible ways a given operator $O_k(\ell+1)$ can act on the first three qubits given the action of $O_k(\ell)$ of the first three qubits. Using Eq.~\eqref{eq:4throw} and the initial condition $O_k(1)=Z_1$, one can show that the action on the first three qubits will never be trivial. Therefore, condition (a) holds, and we have proven the Lemma. \hfill $\Box$

{
We remark that Lemmas \ref{lemma1} and \ref{lemma2} can be turned into a constructive procedure for obtaining any Pauli $P$ as a sequence of commutators of the operators $\widetilde{O}_k(\ell)$. We now describe this procedure and its complexity as a function of $k$.
There are three different classes of $P$ to consider. The first class contains elements of the form $P=Q\otimes I$, where here and in the rest of this discussion, the tensor product separates the first three logical qubits from the rest. Such Paulis can be generated by simply taking commutators of elements in $S_1$. The maximum number of commutators needed to generate an arbitrary element, and the complexity of finding the correct sequence of commutators, do not scale with $k$. Therefore, we do not need to consider the time taken to generate operators of the form $Q\otimes I$ at any step. The second class of Paulis is those of the form $P = Q_1\otimes Q_2$ with $Q_1,Q_2\neq I$. To generate such operators, we need to identify the elements of $S_2$ whose product gives $Q_1'\otimes Q_2$ for some $Q_1'\neq I$ (we comment on how to identify these elements below).  The length of this product is at worst the size of $S_2$, which grows linearly in $k$. The proof of Lemma \ref{lemma1} shows how we can effectively obtain each product of two Paulis by using a constant number of commutators assisted by elements of the form $Q\otimes I$, which can be generated in constant time as stated above. Therefore, the length of the sequence of commutators needed to generate any Pauli of the form $Q_1\otimes Q_2$ grows at most linearly with $k$. Finally, the third case is operators of the form $P=I\otimes Q$. Lemma \ref{lemma1} shows how to obtain any such Pauli as the commutator of two elements of the second class. Therefore, any Pauli $P$ can be obtained as a sequence of commutators, where the length of the sequence is at most linear in $k$. 

Finally, we discuss the complexity of obtaining this sequence. The most costly step is where we identify how to decompose $Q_1'\otimes Q_2$ as a product of elements of $S_2$ for some $Q_1'$. This problem can be cast as a simple linear system of equations and solved via standard Gaussian elimination, which has complexity no worse than $O(k^3)$ since there are $O(k)$ elements in $S_2$. All other steps, such as computing $\Lambda(P,Q)$ or finding Paulis of the form $P\otimes I$ satisfying certain properties, are also at most polynomially difficult in $k$. Therefore, the complexity of compiling operators of the form $e^{i\theta P}$ for arbitrary Pauli strings $P$ is polynomial in $k$.
}
}

\vspace{5mm}
\noindent
\textit{Proof of Lemma~\ref{lemma3}.}
This proof is very similar to that given in Ref.~\onlinecite{Stephen2019}, but some modification is needed due to the open boundary conditions. To account for the open boundary conditions, we follow Ref.~\onlinecite{Raussendorf2005}. Consider the circuit on an infinite chain $T_\infty = \lim_{k\rightarrow \infty} T_k$ with sites indexed by an integer $i\in\mathbb{Z}$. It is shown in Ref.~\onlinecite{Raussendorf2005} that the open boundaries effectively act as a mirror that reflects the information that is propagated out of the range $1\leq i\leq k$ back into that range. This reflection can be mimicked by adding ``ghost images'' of the operator evolution that start outside of this range, similar to the method of image charges from electrostatics. More precisely, the evolution on a finite chain is related to that on an infinite chain by the following equation,
\begin{equation} \label{eq:infinityrelation}
    T_k^\dagger Z_i T_k = T_\infty^\dagger \left[\prod_{n=-\infty}^{\infty} Z_{i+2n(k+1)}Z_{-i+2n(k+1)} \right] T_\infty,
\end{equation}
where operators on the right-hand side acting on qubits outside of the range $1\leq i \leq k$ are simply ignored. A similar condition holds for $T_k^\dagger X_i T_k$, and therefore for any product of Pauli operators. 

We now transition to a more compact notation for describing the action of $T_\infty$, which is translationally invariant~\cite{Schlingemann2008}. As in Section~\ref{sec:byproduct_prop}, let $\xi=(\xi^X;\xi^Z)$ be an infinite binary vector such that Pauli operators on the infinite chain are written, up to a phase, as $V(\xi) =\bigotimes_{i\in \mathbb{Z}} X_i^{\xi^X_i}Z_i^{\xi^Z_i}$. Now, we further condense this information by mapping the binary vector onto a polynomial vector as,
\begin{equation} 
    \begin{pmatrix} {\xi}^X  \\ {\xi}^Z \end{pmatrix} \mapsto \begin{pmatrix} \sum_i (u^i)^{\xi^X_i}  \\ \sum_i (u^i)^{\xi^Z_i} \end{pmatrix}.
\end{equation}
where $u$ is a dummy variable. We denote both the binary and polynomial vectors as $\xi$. Then, we represent the $T_\infty$ as a $2\times 2$ matrix $t_\infty$ of polynomials by arranging 
\begin{equation}
\begin{aligned}
T_\infty(X_0)&=Z_0=\begin{pmatrix} 0 \\ 1\end{pmatrix}:= t_\infty\begin{pmatrix} 1 \\ 0\end{pmatrix},\\
T_\infty(Z_0)&=Z_{-1}Y_0Z_1=\begin{pmatrix} 1 \\ u+1+u^{-1}\end{pmatrix}:= t_\infty\begin{pmatrix} 0 \\ 1\end{pmatrix} 
\end{aligned}
\end{equation} 
into columns of a matrix, giving,
\begin{equation}
t_\infty=\begin{pmatrix} 0 & 1 \\ 1 & u+1+u^{-1} \end{pmatrix}.
\end{equation}
Given a Pauli operator represented by a polynomial vector $\xi$, Eq.~\eqref{eq:infinityrelation} tells us to look at the action of $t_\infty$ on the polynomial $\tilde{\xi}=q_k(u)(\xi + \bar{\xi})$ where $\bar{\xi}$ is obtained from $\xi$ by inverting every monomial $u^i\mapsto u^{-i}$ and $q_k(u)=\sum_{n=-\infty}^{\infty} u^{2n(k+1)}$. We want to show that, if $k=2^r-1$, then 
\begin{equation} \label{eq:tinftyrel}
t_\infty^{3k+3}\tilde{\xi}=(t_\infty^3)^{2^r}\tilde{\xi}=\tilde{\xi}.
\end{equation} Translating back to operator notation, this result, combined with Eq.~\eqref{eq:infinityrelation}, would give $T_k^{3k+3}=I$. 

Assume that $k=2^r-1$. To show that Eq.~\eqref{eq:tinftyrel} is true, we use the Cayley-Hamilton theorem, which says that \cite{Guetschow2010},
\begin{equation} \label{eq:cayley}
(t_\infty^3)^2=\mathrm{Tr}(t_\infty^3) t_\infty^3 + I.
\end{equation}
Let us denote $\gamma = \mathrm{Tr}(t_\infty^3)$. Another important result is that, for any polynomial $p(u)$, we have $p(u)^{2^n} \equiv p(u^{2^n})$ for any $n\in \mathbb{N}$. This holds because the coefficients of the polynomials are in $\mathbb{Z}_2$, so the cross terms cancel out each time $p(u)$ is squared. Combining this with Eq.~\eqref{eq:cayley}, we have,
\begin{equation} \label{eq:t2r}
\begin{aligned}
    (t_\infty^3)^{2^r} &= ((t_\infty^3)^2)^{2^{r-1}} = (\gamma t_\infty^3 + I)^{2^{r-1}} \\
    &= \gamma^{2^{r-1}} (t_\infty^3)^{2^{r-1}} + I.
\end{aligned}
\end{equation}
It is straightforward to compute that,
\begin{equation}
    \gamma = (u^{1}+1+u^{-1})(u^2+u^{-2}).
\end{equation}
This gives,
\begin{equation}
    \gamma^{2^{r-1}} q_k(u) = (u^{2^{r-1}}+1+u^{-2^{r-1}})(u^{2^r}+u^{-2^r}) q_k(u).
\end{equation}
As $q_k(u)=\sum_{n=-\infty}^{\infty} u^{n2^{r+1}}$, we have $(u^{2^r}+u^{-2^r}) q_k(u)=0$ since the two terms are equal and the polynomials have $\mathbb{Z}_2$ coefficients. This means that when computing $(t_\infty^3)^{2^r}\tilde{\xi}$, the first term in Eq.~\eqref{eq:t2r} is equal to zero, so we have $(t_\infty^3)^{2^r}\tilde{\xi} =\tilde{\xi}$ as desired. This completes the proof. \hfill $\Box$

\section{SPT order in the resource states} \label{sec:spt}

In this section we derive the symmetries and SPT order of the states $|\psi_k\rangle$. To do this, we first define the states $|\widetilde{\psi}_k\rangle$ in the same way as $|\psi_k\rangle$ except with periodic boundary conditions, such that qubits 1 and $N$ interact with a $CZ$ each layer of the circuit. Let us assume for simplicity that the length $N$ is a multiple of the period $p_k$. The generalization of Eq.~\eqref{eq:qca_rules} for periodic boundary conditions, taking $\theta_i=0$ for all $i$, is,
\begin{equation}
    \langle s_1,\dots, s_N|\widetilde{\psi}_k\rangle = \mathrm{Tr}\left( A(s_N)\dots A(s_1)\right)
\end{equation}
where $A(s)=U(0,s)=T_kZ_1^s$. This defines a translationally-invariant matrix product state (MPS) representation of the state $|\psi'_k\rangle$~\cite{Cirac2021}. As is generally the case for MPS, the symmetries of the state $|\widetilde{\psi}_k\rangle$ can be understood via the symmetries of the matrices $A(s)$. Using the binary vector notation for Pauli operators introduced in Section~\ref{sec:byproduct_prop}, $A(s)$ satisfies the following relation,
\begin{equation} \label{eq:tensor_symm}
    V(t_k\xi)^\dagger A(s) V(\xi) = (-1)^{s\xi^X_1} A(s).
\end{equation}
This relation can be concatenated $N$ times to derive a symmetry of the wavefunction,
\begin{equation}
\begin{aligned}
    &\mathrm{Tr}\left( A(s_N)\dots A(s_1) \right) \\ &= \left[\prod_{i=1}^N (-1)^{s_i (t_k^i\xi)^X_1}\right]\mathrm{Tr}\left( A(s_N)\dots A(s_1) \right)
\end{aligned}
\end{equation}
which holds for any $\xi$, where $(t_k^{i}\xi)^X_1$ again denotes the first entry of the $X$-component of the vector $t_k^i\xi$. This translates to a symmetry $|\psi_k'\rangle = u(\xi)|\psi_k'\rangle$ where,
\begin{equation} \label{eq:symmetry}
    u(\xi) = \prod_{i=1}^N Z_i^{(t_k^i\xi)^X_1}
\end{equation}
For example, the symmetry $u(\hat{Z_1})$, where $\hat{Z_1}=(0,\dots,0;1,0,\dots,0)$ is the vector representing the Pauli $Z_1$, is reflected in the top row of the space-time diagrams depicting the space-time evolution of the operators $O_k(\ell)$, see Fig.~\ref{fig:qca_evo}. At the end of this section, we prove that all $2^{2k}$ choices of $\xi$ lead to a different symmetry operator when $N$ is a multiple of $p_k$, so the representation $\xi\mapsto u(\xi)$ defines a $\mathbb{Z}_2^{2k}$ symmetry group of $|\widetilde{\psi}_k\rangle$. We also prove that this symmetry group can be fully generated by products of $u(\hat{Z_1})$ and its translations. Note that, if $N$ is not a multiple of $p_k$, the global symmetry group will be smaller, but the bulk symmetry, and therefore the SPT order, is not affected.

Having identified the symmetries of the resource states, we now ask whether they have non-trivial SPT order with respect to these symmetries. Our calculations used to derive the symmetry contain enough information to answer this question. Namely, Eq.~\eqref{eq:tensor_symm} reveals that the representation of the symmetry acting on the virtual Hilbert space is $\xi\mapsto V(\xi)$. While the physical representation $\xi\mapsto u(\xi)$ is \textit{linear}, meaning $u(\xi)u(\zeta)=u(\xi\oplus \zeta)$ where $\oplus$ denotes mod-2 addition, the virtual representation is \textit{projective} with,
\begin{equation} \label{eq:symp}
    V(\xi)V(\zeta)=(-1)^{\xi^T\Lambda\zeta}V(\xi\oplus\zeta)
\end{equation}
where $T$ denotes the transpose and $\Lambda = \begin{pmatrix} 0 & I_k \\ I_k & 0 \end{pmatrix}$ with $I_k$ the $k\times k$ identity matrix. The non-commutativity of the virtual representation of the symmetry implies the existence of non-trivial SPT order~\cite{Pollmann2010,Chen2011,Schuch2011}. In fact, this representation is a so-called maximally non-commutative representation, which is one where only the identity element commutes with all other elements~\cite{Else2012}. Using the results of~\cite{Else2012}, it is straightforward to show that this implies that the degeneracy of the zero-energy edge mode associated to the SPT phase is $\sqrt{|\mathbb{Z}_2^{2k}|}=2^{k}$, since this is the minimal dimension of irreducible representation satisfying Eq.~\eqref{eq:symp}. Maximally non-commutative phases are also exactly the phases to which the general results of Refs.~\onlinecite{Else2012,Stephen2017,Stephen2019} apply. Using these results, we can extend our MBQC protocol beyond the fixed-point states $|\psi_k\rangle$.

Finally, we prove two useful results about the symmetry representation $\xi\mapsto u(\xi)$ defined in Eq.~\eqref{eq:symmetry}. First we show that this is a faithful representation of $\mathbb{Z}_2^{2k}$. That is, if $u(\xi)=I$, then $\xi=0$. By definition, if $u(\xi)=I$, then $(t_k^i\xi)^X_1=0$ for all $i$. Since $T_k^\dagger X_1T_k=Z_1$, we have $(t_k^i\xi)^X_1=(t_k^{i+1}\xi)^Z_1$. So, if $(t_k^i\xi)^X_1=0$ for all $i$, then we also have $(t_k^i\xi)^Z_1=0$ for all $i$. So, in the operator evolution of $V(\xi)$ under $T_k$, the first row corresponding to $i=1$ remains empty (\textit{i.e.} acted on by $I$) for all times. One can quickly convince oneself that, given the rules in Eq.~\eqref{eq:qca_rules}, this is only possible if $\xi=0$. For example, if the first row is empty, then the second row must contain no $Z$ operators, since a $Z$ operator in the second row would spread to the first row in the next column, and therefore must also contain no $X$ operators. Progressing in this way, we conclude that every row must be empty, so $\xi=0$. Therefore, $u(\xi)=0$ implies $\xi=0$, so the representation is faithful.

Now, we prove that $u(\xi)$ can be expressed as a product of translations $u(\hat{Z_1})$ for all $\xi$. Translations of $u(\hat{Z_1})$ can be written as $u(t_k^\ell\hat{Z_1})$ for $\ell=0,\dots, p_k$. We want to prove that every vector $\xi$ can be expressed as a sum of the vectors $t_k^\ell\hat{Z_1}$. If this is true, then $u(\xi)$ is always a product of $u(t_k^\ell\hat{Z_1})$ for different $\ell$, since $u(\xi)u(\zeta)=u(\xi\oplus\zeta)$. The required condition is equivalent to the fact that the operators $O_k(\ell)\equiv T_k^{\ell\dagger} Z_1 T_k^{\ell}$ generate the whole set of Pauli operators under multiplication. This is in turn equivalent to saying that the set $\{A(s_1)\dots A(s_L):s_1,\dots,s_L=0,1\}$ for $L=p_k$ spans the whole space of $2^k\times 2^k$ matrices, where $A(s)$ are the MPS tensors defined above. This property of an MPS is known as injectivity. Physically, an MPS has finite correlation length if and only if it admits an injective MPS representation~\cite{Cirac2021}. The states $|\psi'_k\rangle$ certainly have finite correlation length, as they are generated from a product state with a quantum circuit, which implies that there exists an MPS representation of $|\psi'_k\rangle$ given by some matrices $B(s)$ which satisfy injectivity. To prove that our matrices $A(s)$ are also injective, we first observe that $|\psi'_k\rangle$ has a half-chain entanglement entropy of $S=k\log 2$, which is in fact a consequence of dual-unitarity~\cite{Zhou2022}. For an MPS defined by matrices of dimension $D$, the half-chain entanglement entropy is upper bounded by $\log D$. This means that $A(s)$, which has dimension $2^k$, has the minimal possible bond dimension. Therefore, by Proposition 20 of Ref.~\onlinecite{Molnar2018}, there exists an invertible matrix $X$ such that $A(s_1)\dots A(s_n)=XB(s_1)\dots B(s_n)X^{-1}$ for large enough $n$, so the matrices $A(s)$ must also satisfy injectivity. This proves that $O_k(\ell)$ generate all Pauli operators under multiplication, which proves our desired result on $u(\xi)$.

\section{A non-universal dual-unitary Clifford circuit} \label{sec:clusterqca}

\begin{figure}
    \includegraphics[width=\linewidth]{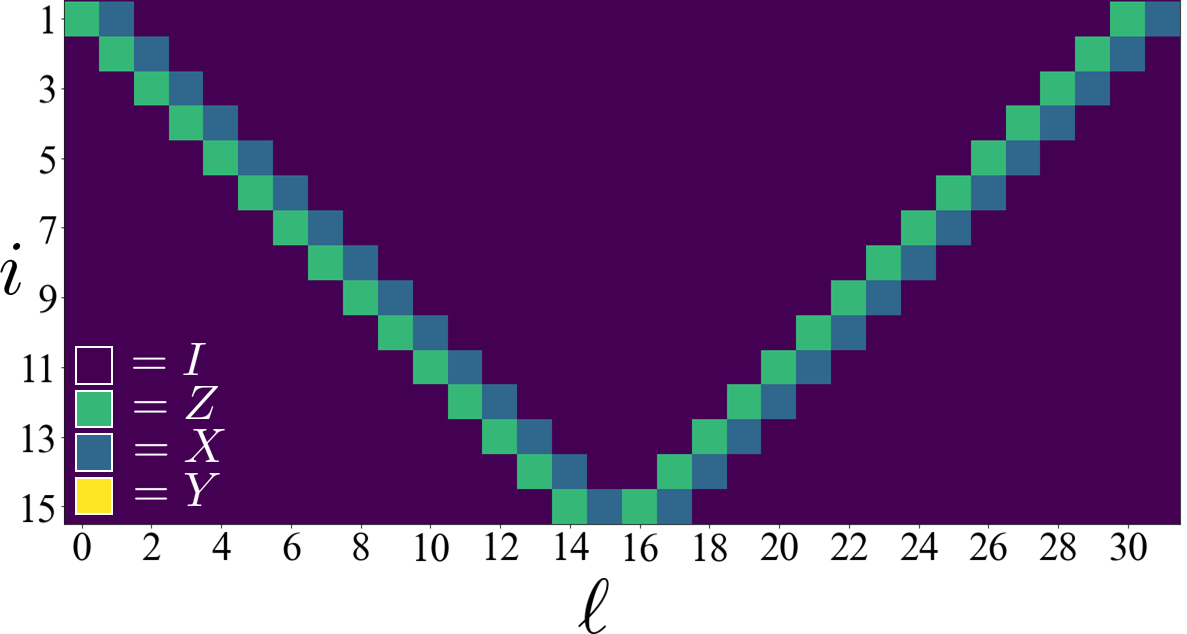}
    \caption{(a) The operator evolution inducted by ${T}'_k$ which is obtained from $T_k$ by replacing $R\mapsto H$, with $k=15$. Each column represents an operator $O'_k(\ell)$.}
    \label{fig:clus_qca}
\end{figure}

In this section we give an example of an alternate family of resource states which are also generated by a dual unitary circuit. They also share the feature that they are fixed-points of certain SPT phases. However, they differ in the fact that the set of implementable gates when using them as resource states for MBQC is not universal.

These states are defined as follows,
\begin{equation} \label{eq:psitildek}
    |\psi'_k\rangle = T_N^{\prime k} \left( \bigotimes_{i=1}^N |+\rangle_i\right),
\end{equation}
where,
\begin{equation} \label{eq:qcatilde}
   T'_N = \prod_{i=1}^{N}H_i \prod_{i=1}^{N-1} CZ_{i,i+1}.
\end{equation}
These states differ from the states defined in the main text only by the substitution $HS \rightarrow H$. This corresponds to choosing $g=0$ in the kicked-Ising Hamiltonian. It is straightforward to show that most of the claims and constructions of the main text follow through with $T$ replaced by $T'$. In particular, the circuit is still dual unitary such that viewing it sideways reveals a quantum computation in virtual space that is controlled by the measurement bases, where the set of unitary gates that can be performed are again rotations generated by operators of the form ${T'_k}^{\dagger} Z_1T'_k$. {One can also show that the states $|\psi'_k\rangle$ again possess SPT order with symmetry $\mathbb{Z}_2^{2k}$}.

The difference compared to the case presented in the main text is that these rotations no longer form a universal set. This is a consequence of the much simpler space-time evolution of the operators under conjugation by $T'_k$, \textit{i.e.} the operators $O'_k(\ell)={T'_k}^{\ell\dagger}Z_1{T'_k}^{\ell}$, see Fig.~\ref{fig:clus_qca}. We see that $O'_k(\ell)$ always has a basic form of a two-qubit operator (except at the boundaries) that is then translated along the virtual Hilbert space upon conjugation by $T'_k$. This is a consequence of the fact that, if we regard $T'_k$ as a Clifford quantum cellular automaton (CQCA), then it falls into the ``glider'' class as defined in Ref.~\onlinecite{Guetschow2010} (see also Ref.~\onlinecite{Stephen2019}). This class of CQCA is defined by the existence of operators on which the CQCA acts as translation. On the other hand, if we regard $T_k$ as a CQCA, it falls into the ``fractal'' class~\cite{Guetschow2010}, which is less fine-tuned and has more complicated dynamics, as we can readily observe in our context. 

As a consequence of the simple form of the operators $O'_k(\ell)$, the algebra $\mathcal{A}'_k$ that they generate under commutation is also simple. The size of this algebra turns out to grow quadratically in $k$, as we have $dim(\mathcal{A}_k)=(k+1)(2k+1)$, which matches the dimension of $so(2(k+1))$. Such a polynomially growing algebra is not sufficient for universal MBQC (which requires the algebra to grow exponentially in $k$)~\cite{Somma2006}.

To see the non-universaility more explicitly, we show that the MBQC protocol using the state $|\psi'_k\rangle$ can be efficiently simulated with a classical computer. Given the simple form of $O'_k(\ell)$ displayed in Fig.~\ref{fig:clus_qca}, one can show that the elements of the algebra $\mathcal{A}'_k$ each have a form that falls into one of three classes: $Y_1\dots Y_{i-1}U_i$, $U_i Y_{i+1}\dots Y_k$, or $U_i Y_{i+1}\dots Y_{j-1} V_{j}$ for $j\geq i$ where $U,V=X$ or $Z$. Let us modify the first two classes of operators by adding two new qubits at the ends of the chain, indexed by $i=0,k+1$, and defining corresponding operators, respectively, as $X_0Y_1\dots Y_{i-1}U_i$, $U_i Y_{i+1}\dots Y_k X_{k+1}$. Then, all three classes have the form $U_i Y_{i+1}\dots Y_{j-1} V_{j}$ where $0\leq i\leq j \leq k+1$. These operators correspond to quadratic fermionic operators after a Jordan-Wigner transformation. Therefore the expectation value,
\begin{equation}
        \langle 0^{\theta_1},\dots,0^{\theta_N}|\psi'_k\rangle = \langle R|e^{i\theta_N O'_k(N)}\dots e^{i\theta_1 O'_k(1)}|L\rangle
\end{equation}
where $|L\rangle$ is as defined before and $\langle R|$ can be any product state in the $Z$-basis, can be efficiently calculated using techniques from matchgate circuits (equivalently, fermionic linear optics)~\cite{Valiant2002,Terhal2002,Jozsa2008}. Therefore, the MBQC protocol can be efficiently classically simulated.

\begin{figure*}[h]
	\subfigure[ $\lambda_{i,t}=1$.]{{\label{fig:duc1}\begin{tabular}[b]{|c|cccccc|}\hline
      $k$ & 2 & 3 & 4 & 5 & 6 & 7  \\ \hline\hline
      $p_k$ & 4 & 12 & 10 & 24 & 18 & 24  \\ \hline
      dim($\mathcal{A}_k$) & 10 & 63 & 120 & 496 & 4095 & 8256  \\ \hline
      $\mathcal{A}_k$  &  $sp(2^k)$ & $su(2^k)$ & $so(2^k)$ & $so(2^k)$ & $su(2^k)$ & $sp(2^k)$  \\ \hline
    \end{tabular}}}
    \hspace{1cm}
	\subfigure[$\lambda_{i,t}=0$.  $\left(k'=2(k+1)\right)$]{{\label{fig:duc2}\begin{tabular}[b]{|c|cccccc|}\hline
      $k$ & 2 & 3 & 4 & 5 & 6 & 7  \\ \hline\hline
      $p_k$ & 6 & 8 & 10 & 12 & 14 & 16  \\ \hline
      dim($\mathcal{A}_k$) & 15 & 28 & 45 & 66 & 91 & 120  \\ \hline
      $\mathcal{A}_k$  & $so(k')$ & $so(k')$ & $so(k')$ & $so(k')$ & $so(k')$ & $so(k')$  \\ \hline
    \end{tabular}}}
	\subfigure[$\lambda_{i,t}=\delta({k-t\equiv 1}\mod 2$).]{{\label{fig:duc3}\begin{tabular}[b]{|c|cccccc|}\hline
      $k$ & 2 & 3 & 4 & 5 & 6 & 7  \\ \hline\hline
      $p_k$ & 5 & 12 & 17 & 10 & 63 & 24  \\ \hline
      dim($\mathcal{A}_k$) & 10 & 36 & 255 & 496 & 2016 & 8128  \\ \hline
      $\mathcal{A}_k$  & $sp(2^k)$ & $sp(2^k)$ & $su(2^k)$ & $so(2^k)$ & $so(2^k)$ & $so(2^k)$  \\ \hline
    \end{tabular}}}
    \hspace{1cm}
	\subfigure[$\lambda_{i,t}=\delta({k-t\equiv 1}\mod 4$).]{{\label{fig:duc4}\begin{tabular}[b]{|c|cccccc|}\hline
      $k$ & 2 & 3 & 4 & 5 & 6 & 7  \\ \hline\hline
      $p_k$  &5 & 12 & 17 & 30 & 63 & 48  \\ \hline
      dim($\mathcal{A}_k$) & 10 & 36 & 136 & 528 & 4095 & 8128  \\ \hline
      $\mathcal{A}_k$  & $sp(2^k)$ & $sp(2^k)$ & $sp(2^k)$ & $sp(2^k)$ & $su(2^k)$ & $so(2^k)$  \\ \hline
    \end{tabular}}}
	\subfigure[$\lambda_{i,t}=\delta({i\equiv 1}\mod 2$).]{{\label{fig:duc5}\begin{tabular}[b]{|c|cccccc|}\hline
      $k$ & 2 & 3 & 4 & 5 & 6 & 7  \\ \hline\hline
      $p_k$ & 8 & 16 & 12 & 16 & 36 & 32  \\ \hline
      dim($\mathcal{A}_k$) & 15 & 63 & 255 & 1023 & 4095 & 16383  \\ \hline
      $\mathcal{A}_k$  & $su(2^k)$ & $su(2^k)$ & $su(2^k)$ & $su(2^k)$ & $su(2^k)$ & $su(2^k)$  \\ \hline
    \end{tabular}}}
    \hspace{1cm}
	\subfigure[$\lambda_{i,t}=\delta({i\equiv 1}\mod 4$).]{{\label{fig:duc6}\begin{tabular}[b]{|c|cccccc|}\hline
      $k$& 2 & 3 & 4 & 5 & 6 & 7  \\ \hline\hline
      $p_k$ &16 & 8 & 40 & 32 & 56 & 16  \\ \hline
      dim($\mathcal{A}_k$) & 15 & 28 & 255 & 1023 & 4095 & 16383  \\ \hline
      $\mathcal{A}_k$  & $su(2^k)$ & $so(2^k)$ & $su(2^k)$ & $su(2^k)$ & $su(2^k)$ & $su(2^k)$  \\ \hline
    \end{tabular}}}
	\subfigure[$\lambda_{i,t}=\delta({i\equiv 1}\mod 16$).]{{\label{fig:duc7}\begin{tabular}[b]{|c|cccccc|}\hline
      $k$ & 2 & 3 & 4 & 5 & 6 & 7  \\ \hline\hline
      $p_k$ & 52 & 24 & 150 & 116 & 274 & 32  \\ \hline
      dim($\mathcal{A}_k$) & 15 & 28 & 255 & 1023 & 4095 & 120  \\ \hline
      $\mathcal{A}_k$  & $su(2^k)$ & $so(2^k)$ & $su(2^k)$ & $su(2^k)$ & $su(2^k)$ & $so(2(k+1))$  \\ \hline
    \end{tabular}}}
    \hspace{1cm}
	\subfigure[$\lambda_{i,t}=\delta({i\equiv 1}\mod 2\text{ and } k-t\equiv 0\mod 2)$.]{{\label{fig:duc8}\begin{tabular}[b]{|c|cccccc|}\hline
      $k$& 2 & 3 & 4 & 5 & 6 & 7  \\ \hline\hline
      $p_k$ &12 & 8 & 20 & 12 & 28 & 16  \\ \hline
      dim($\mathcal{A}_k$) & 15 & 21 & 255 & 528 & 4095 & 8128  \\ \hline
      $\mathcal{A}_k$  & $su(2^k)$ & $sp(2k)$ & $su(2^k)$ & $sp(2^k)$ & $su(2^k)$ & $so(2^k)$  \\ \hline
    \end{tabular}}}
    \subfigure[$\lambda_{i,t}=\delta(t=2)$.]{{\label{fig:duc9}\begin{tabular}[b]{|c|cccccc|}\hline
      $k$ & 2 & 3 & 4 & 5 & 6 & 7  \\ \hline\hline
      $p_k$ & 5 & 12 & 17 & 30 & 93 & 180  \\ \hline
      dim($\mathcal{A}_k$) & 15 & 36 & 255 & 496 & 4095 & 8256  \\ \hline
      $\mathcal{A}_k$  & $su(2^k)$ & $sp(2^k)$ & $su(2^k)$ & $so(2^k)$ & $su(2^k)$ & $sp(2^k)$  \\ \hline
    \end{tabular}}}
    \hspace{1cm}
	\subfigure[$\lambda_{i,t}=\delta(t=1)$.]{{\label{fig:duc10}\begin{tabular}[b]{|c|cccccc|}\hline
      $k$& 2 & 3 & 4 & 5 & 6 & 7  \\ \hline\hline
      $p_k$ &5 & 7 & 9 & 11 & 13 & 15  \\ \hline
      dim($\mathcal{A}_k$) & 10 & 21 & 36 & 55 & 78 & 105  \\ \hline
      $\mathcal{A}_k$  & $sp(2k)$ & $sp(2k)$ & $sp(2k)$ & $sp(2k)$ & $sp(2k)$ & $sp(2k)$  \\ \hline
    \end{tabular}}}
    \caption{Numerical calculations of computational power at small $k$ for different choices of the matrix $\lambda_{i,t}$. The matrices are defined in terms of functions $\delta$ which return 1 if the contained expression is true and 0 otherwise. Tables~\subref{fig:duc1} and~\subref{fig:duc2} correspond to the cases considered in the main text and in Section~\ref{sec:clusterqca} and are given for the sake of comparison. In table~\subref{fig:duc2}, we write  $k'=2(k+1)$ for compactness. Observe that all tables except~\subref{fig:duc2} and~\subref{fig:duc10} show that $\mathcal{A}_k$ appears to grow exponentially with $k$. In Table~\subref{fig:duc7}, the non-universality at $k=7$ is an artifact of a special relation between $k$ and the periodicity of $\lambda$; this case goes back to being universal at $k=8$.}
    \label{fig:tables}
\end{figure*}

\begin{figure*}[h]
    \centering
    \includegraphics[width=0.7\linewidth]{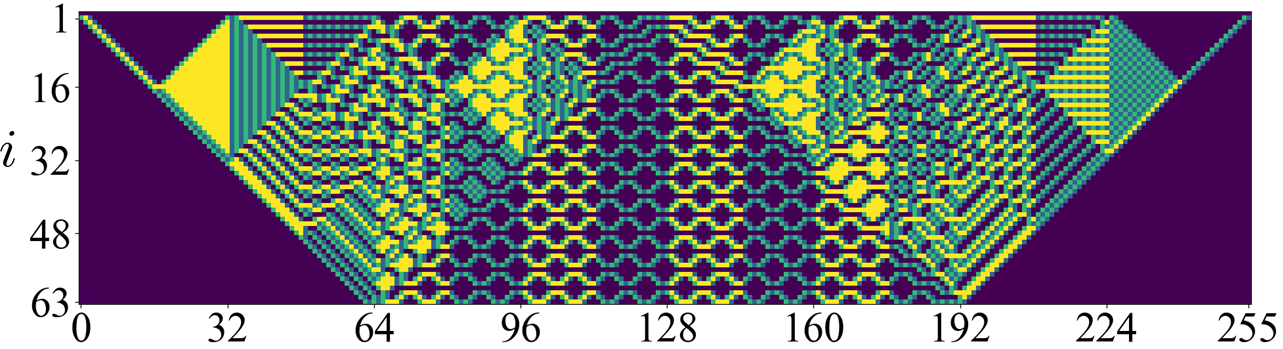}\hfill
    \includegraphics[width=0.7\linewidth]{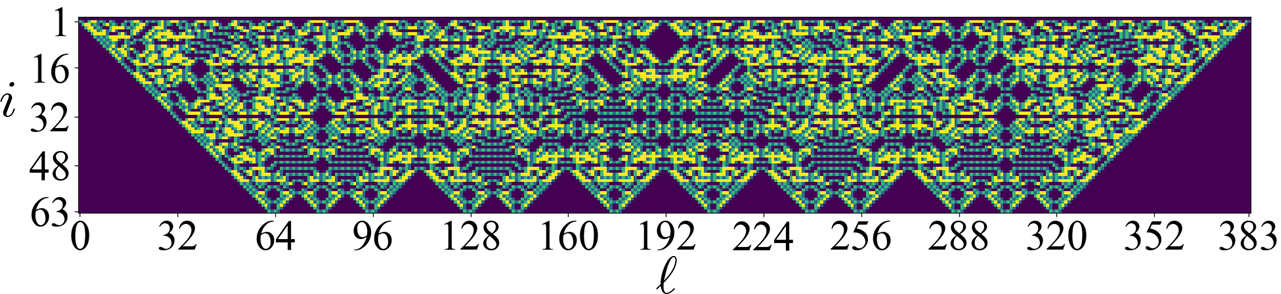}
    \caption{Spacetime evolution of the $Z_1$ for the $k=63$ instance of the states defined by $\lambda_{i,t}$ as given in Table \ref{fig:duc7} (upper) and Table \ref{fig:duc4} (lower).}
    \label{fig:qca_evos_app}
\end{figure*}

\section{More examples of dual-unitary Clifford circuits} \label{sec:cliffs}

In this section, we investigate more examples of dual-unitary Clifford (DUC) circuits which increase the size of the unit cell for translation invariance in space or time. The goal of this study is to support the claim that the non-universality of the dual-unitary circuit studied in Section \ref{sec:clusterqca}---as indicated by the polynomially-sized Lie algebra of generators of rotations---is fine-tuned, and that generic DUC circuits will have exponentially growing Lie algebras. To construct these examples, we consider the same basic gates $CZ$, $H$, and $HS$. At each time step, we apply $CZ$ to all nearest neighbours, followed by either $H$ or $HS$ on each qubit depending on both the spatial location of the qubit and the time step. 

More precisely, we construct the following resource states,
\begin{equation}
    |\psi_k(\lambda)\rangle = \left(\prod_{t=1}^k T_{N,t}(\lambda) \right) \left( \bigotimes_{i=1}^N |+\rangle_i\right)
\end{equation}
where,
\begin{equation}
    T_{N,t}(\lambda) = \prod_{i=1}^N H_i S_i^{\lambda_{i,t}}\prod_{i=1}^N CZ_{i,i+1}.
\end{equation}
The matrix elements $\lambda_{i,t}=0,1$ determine whether or not $S$ acts on site $i$ at time-step $t$. For every choice of $\lambda$, the circuit defining $|\psi_k(\lambda)\rangle$ corresponds to a DUC circuit. In this notation, states $|\psi_k\rangle$ defined in the main text correspond to $\lambda_{i,t}=1$, while the states $|\psi'_k\rangle$ defined in Section~\ref{sec:clusterqca} correspond to $\lambda_{i,t}=0$. 

We now numerically study a few examples of the above states. In order to be able to calculate a well-defined period of the virtual evolution, we enforce some periodicity on $\lambda_{i,t}$ in the time direction, but allowing a larger unit cell. For example, we could apply $S$ only on even number qubits, or only during even numbered time steps, and so on. The results are given in Table~\ref{fig:tables}. We see that in all cases, except for the case studied in Section~\ref{sec:clusterqca} and one other, the dimension of $\mathcal{A}_k$ appears to grow exponentially with $k$. 
For two of the universal cases, we also show the space-time evolution of the corresponding operators $O_k(\ell)$ in Fig.~\ref{fig:qca_evos_app}. We observe that these cases display complex structures that are in distinct contrast to the simple evolution of the non-universal case shown in Section~\ref{sec:clusterqca}.

The second non-universal case~\subref{fig:duc10} appears when $S$ is applied only on the first time step. This is equivalent to never applying $S$, and instead changing the initial state of the chain from $\bigotimes_{i=1}^N |+\rangle_i$ to $\bigotimes_{i=1}^N |y_+\rangle_i$ where $|y_+\rangle$ is the $+1$ eigenstate of $Y$. The computation that is induced in the spatial direction in this case is again matchgate circuit as in Section~\ref{sec:clusterqca}.
{It is interesting to note that the family of states generated in this case corresponds to the family of SPT-ordered spin chains studied in Ref.~\cite{Verresen2017} which are Jordan-Wigner equivalent to free-fermion chains.} We note that the non-universality of this case also appears to be fine tuned, as indeed applying $S$ only on the second time-step (instead of the first) as in Table~\ref{fig:duc9} appears to give an exponentially growing algebra. This suggests that any small change in the dual-unitary circuit which drives it away from the matchgate case will lead to universality.

Finally, we note that the proof of universality (for the $\lambda_{i,t}=1$ case) given in Section~\ref{sec:universality_proof} works by numerically proving universality for small $k$ and then taking advantage of scale invariance to prove universality for all $k$. Therefore, it is plausible that the universality at small $k$ shown for the examples in Table~\ref{fig:tables} can similarly be used to prove universality for all $k$.

\end{document}